\documentclass[sn-mathphys,Numbered]{sn-jnl}

\usepackage{graphicx}%
\usepackage{multirow}%
\usepackage{amsmath,amssymb,amsfonts}%
\usepackage{amsthm}%
\usepackage{mathrsfs}%
\usepackage[title]{appendix}%
\usepackage{xcolor}%
\usepackage{textcomp}%
\usepackage{manyfoot}%
\usepackage{booktabs}%
\usepackage{algorithm}%
\usepackage{algorithmicx}%
\usepackage{algpseudocode}%
\usepackage{listings}%

\usepackage[mathscr,scaled=1.15]{urwchancal}
\DeclareFontFamily{OT1}{pzc}{}
\DeclareFontShape{OT1}{pzc}{m}{it}%
{<-> s * [1.15] pzcmi7t}{}
\DeclareMathAlphabet{\mathpzc}{OT1}{pzc}{m}{it}

\theoremstyle{thmstyleone}%
\theoremstyle{thmstyletwo}%

\theoremstyle{thmstylethree}%

\begin{document}

\title[Hadron Structure using Continuum Schwinger Function Methods]
{$\,$\\[-7ex]\hspace*{\fill}{\normalsize{\sf\emph{Preprint no}. NJU-INP 072/23}}\\[1ex]
Hadron Structure using Continuum Schwinger Function Methods}


\author*[1,2]{\fnm{Craig D.} \sur{Roberts}%
       $^{\href{https://orcid.org/0000-0002-2937-1361}{\textcolor[rgb]{0.00,1.00,0.00}{\sf ID}},}$}\email{cdroberts@nju.edu.cn}

\affil*[1]{\orgdiv{School of Physics}, \orgname{Nanjing University}, \orgaddress{\city{Nanjing}, \postcode{210093}, \state{Jiangsu}, \country{China}}}

\affil[2]{\orgdiv{Institute for Nonperturbative Physics}, \orgname{Nanjing University}, \orgaddress{\city{Nanjing}, \postcode{210093}, \state{Jiangsu}, \country{China}}}


\abstract{The vast bulk of visible mass emerges from nonperturbative dynamics within quantum chromodynamics (QCD) -- the strong interaction sector of the Standard Model.  The past decade has revealed the three pillars that support this emergent hadron mass (EHM); namely, a nonzero gluon mass-scale, a process-independent effective charge, and dressed-quarks with constituent-like masses.  Theory is now working to expose their manifold and diverse expressions in hadron observables and highlighting the types of measurements that can be made in order to validate the paradigm.
In sketching some of these developments, this discussion stresses the role of EHM in
forming nucleon electroweak structure and the wave functions of excited baryons through the generation of dynamical diquark correlations;
producing and constraining the dilation of the leading-twist pion distribution amplitude;
shaping pion and nucleon parton distribution functions -- valence, glue and sea, including the antisymmetry of antimatter;
and moulding pion and proton charge and mass distributions.
\\[2ex]
Invited contribution to a Special Issue of Few Body Systems:
``Emergence and Structure of Baryons -- Selected Contributions from the International Conference Baryons 2022''
}




\maketitle

\section{Introduction}\label{sec1}
\begin{figure}[h]
\begin{tabular}{ll}
\hspace*{-6ex}\includegraphics[clip, width=0.51\textwidth]{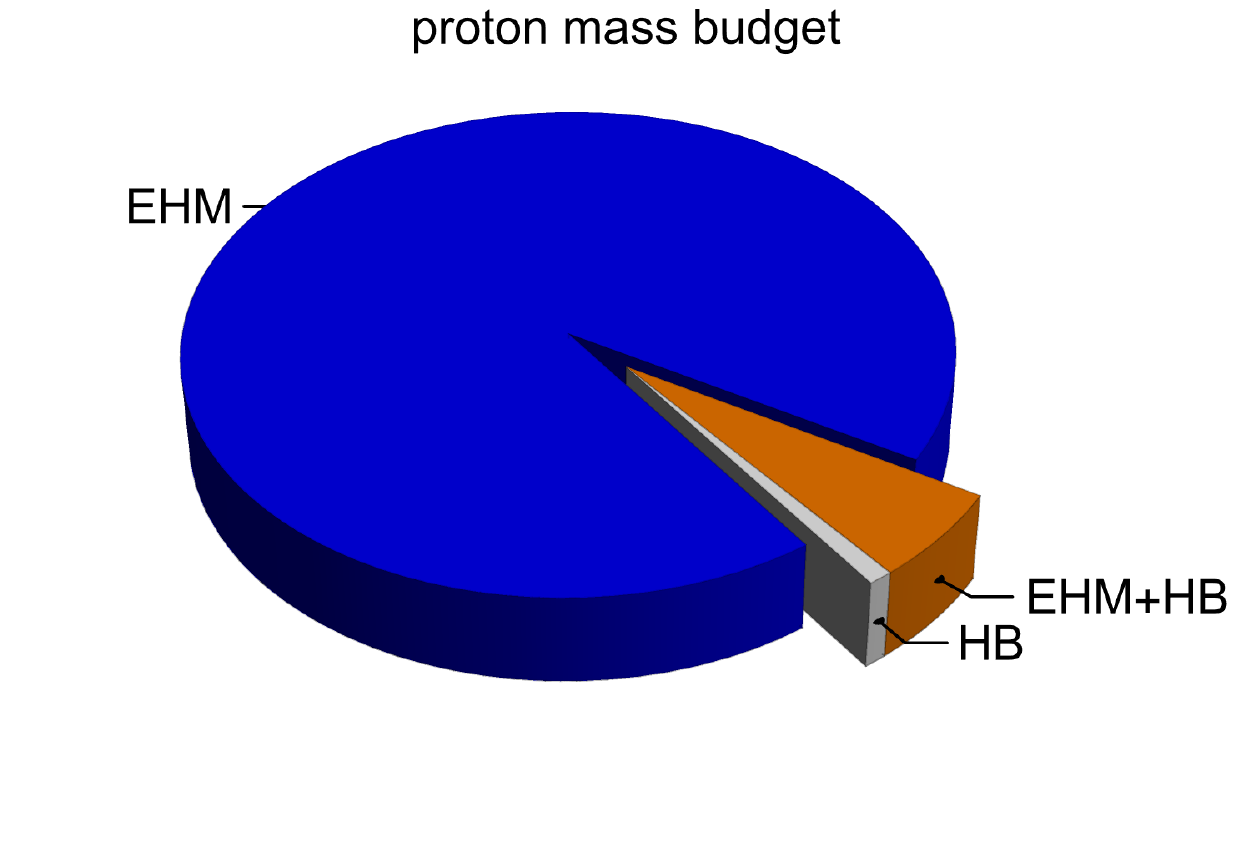}&
\includegraphics[clip, width=0.51\textwidth]{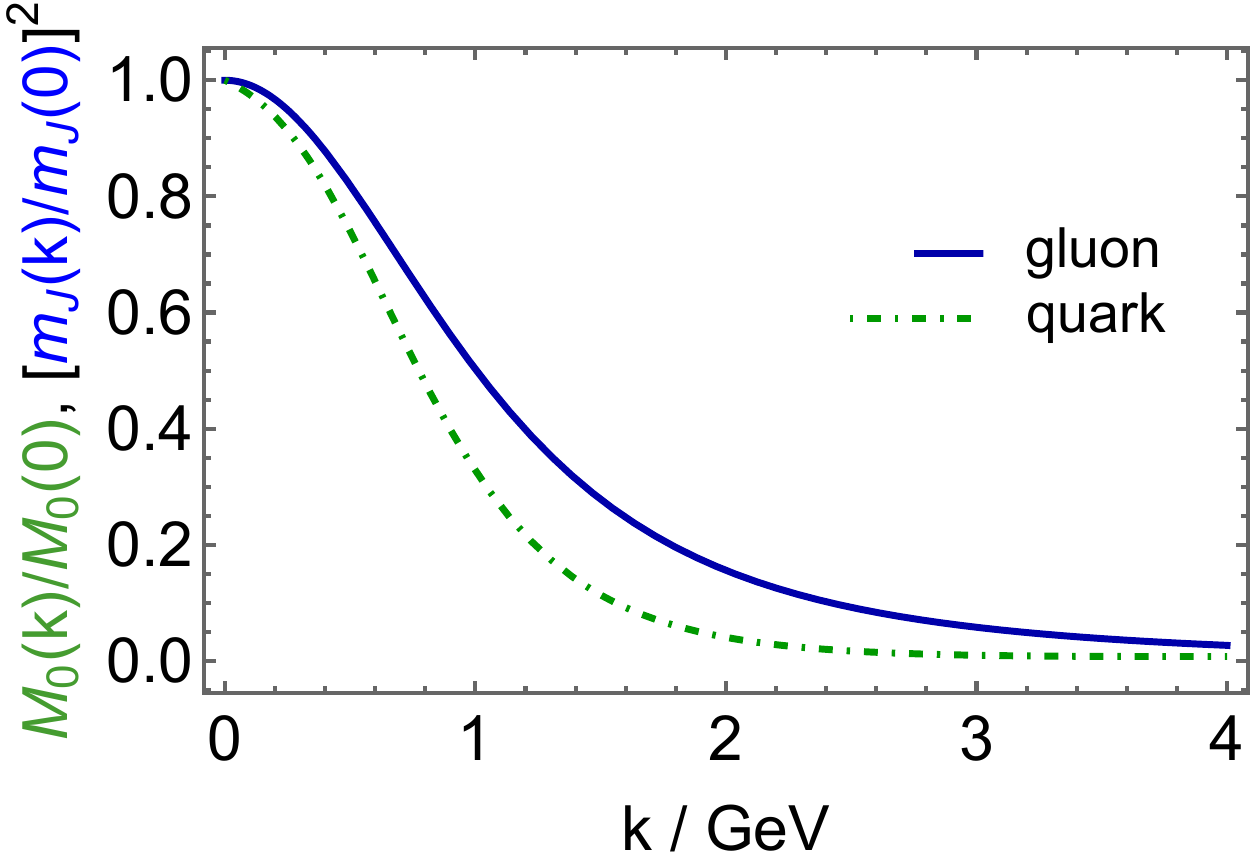} \\[-3ex]
(\textbf{A}) & (\textbf{B})
\end{tabular}
\vspace*{1ex}

\caption{\label{Fmassbudget}
\textbf{A}.
Proton mass budget, drawn using a Poin\-car\'e-invariant decomposition:
EHM $= 94$\%;
Higgs boson (HB) contribution $= 1$\%;
and EHM+HB interference $= 5$\%.
(Separation at renormalisation scale $\zeta = 2\,$GeV, calculated using information from Refs.\,\cite{Flambaum:2005kc, RuizdeElvira:2017stg, Aoki:2019cca, Workman:2022ynf}).
\textbf{B}.
Renormalisation-group-invariant dressed gluon mass function (solid blue curve) calculated, following Ref.\,\cite{Aguilar:2019uob}, from a gluon two-point function obtained using the lattice-QCD configurations in Refs.\,\cite{Blum:2014tka, Boyle:2015exm, Boyle:2017jwu}.  The mass-squared curve is plotted, normalised by its $k=0$ value, and compared with the chiral-limit dressed quark mass function drawn from Ref.\,\cite{Binosi:2016wcx} (dotted-dashed green curve).  This pair of curves is $1/k^2$-suppressed in the ultraviolet, each with additional logarithmic corrections.
}
\end{figure}

The Standard Model of Particle Physics has one widely known mass-generating mechanism; namely, the Higgs boson \cite{Englert:2014zpa, Higgs:2014aqa}.  It is critical to the evolution of our Universe.  Yet, by itself, the Higgs is responsible for only 1\% of the visible mass in the Universe.  Visible matter is constituted from nuclei found on Earth and the mass of each such nucleus is primarily just the sum of the masses of the nucleons they contain.  However, only 9\,MeV of a nucleon's mass, $m_N=940\,$MeV, is directly generated by Higgs boson couplings into quantum chromodynamics (QCD).  Evidently -- as highlighted by Fig.\,\ref{Fmassbudget}A, Nature has another, very effective mass-generating mechanism.
Today, this is called emergent hadron mass (EHM) \cite{Roberts:2020udq, Roberts:2020hiw, Roberts:2021xnz, Roberts:2021nhw, Binosi:2022djx, Roberts:2022rxm, Papavassiliou:2022wrb, Ding:2022ows, Ferreira:2023fva, Carman:2023zke}: it is responsible for 94\% of $m_N$, with the remaining 5\% generated by constructive interference between EHM and the Higgs-boson.

Given the mass budget in Fig.\,\ref{Fmassbudget}A, one is immediately driven to ask the following questions.
What is the dynamical origin of EHM; what are its connections with confinement, itself seemingly characterised by a single nonperturbative mass scale \cite[Sec.\,5]{Ding:2022ows}; and are these phenomena linked with, or even the underlying cause of, dynamical chiral symmetry breaking (DCSB)?
This last, \textit{viz}.\ DCSB, has long been argued to provide the key to understanding the pion, Nature's most fundamental Nambu-Goldstone boson, with its unusually low mass and structural peculiarities \cite{Horn:2016rip, Aguilar:2019teb, Chen:2020ijn, Anderle:2021wcy, Arrington:2021biu, Quintans:2022utc}.

To answer the first question, one must consider sixty-years of work in studying nonperturbative gauge-sector dynamics in Poincar\'e-invariant quantum field theories \cite{Schwinger:1962tn, Schwinger:1962tp, Cornwall:1981zr, Mandula:1987rh}.  Those analyses have culminated in the realisation that a Schwinger mechanism is active in QCD \cite{Binosi:2022djx, Ferreira:2023fva}.  The gluon vacuum polarisation tensor remains four transverse:
\begin{equation}
\Pi_{\mu\nu}(k) = (k^2 \delta_{\mu\nu} - k_\mu k_\nu) \Pi(k^2)\,;
\end{equation}
but $k^2 \Pi(k^2) \neq 0$.  This feature owes principally to gluon self-interactions and entails that the three four-transverse modes of the gluon acquire a momentum dependent mass.  The mass function is power-law suppressed in the ultraviolet, hence, invisible in perturbation theory; yet it is large at infrared momenta, being characterised by a renormalisation-point-independent value \cite{Binosi:2016nme, Cui:2019dwv}:
\begin{equation}
\label{gluonmass}
m_0 = 0.43(1)\,{\rm GeV}.
\end{equation}
The renormalisation-group-invariant (RGI) gluon mass function is drawn in Fig.\,\ref{Fmassbudget}B.

Qualitatively, the result in Eq.\,\eqref{gluonmass} is the definitive statement of \emph{mass emerging from nothing}; thus, the most basic expression of EHM in Nature.  Impossible in perturbative-QCD, Eq.\,\eqref{gluonmass} owes to the synergistic action of infinitely-many massless gluon partons, which fuse and behave together as a coherent quasiparticle gluon with a long-wavelength mass.  That mass is $\approx \tfrac{1}{2} m_N$ and its emergence has far-reaching consequences, some of which are sketched below.

The archetypal running coupling in quantum gauge field theories is the Gell-Mann--Low effective charge in quantum electrodynamics, which features in textbooks -- see, e.g., Ref.\,\cite[Ch.\,13.1]{IZ80FBS}.  It is a RGI, process-independent (PI) running coupling and, owing to the Ward identity \cite{Ward:1950xp, Takahashi:1957xn}, is obtained directly by computing the photon vacuum polarisation.
The presence of ghost fields in standard Poincar\'e-covariant treatments of QCD means that developing an analogue is not straightforward.  Over time, this impediment has spawned many attempts to define a nonperturbative extension of the QCD running coupling using one or another of QCD's various interaction vertices \cite[Sec.\,4.6.3]{Deur:2023dzc}.  These schemes yield what one might call vertex-dependent (VD) effective charges.

\begin{figure}[t]
\hspace*{-1ex}\includegraphics[clip, width=0.54\textwidth]{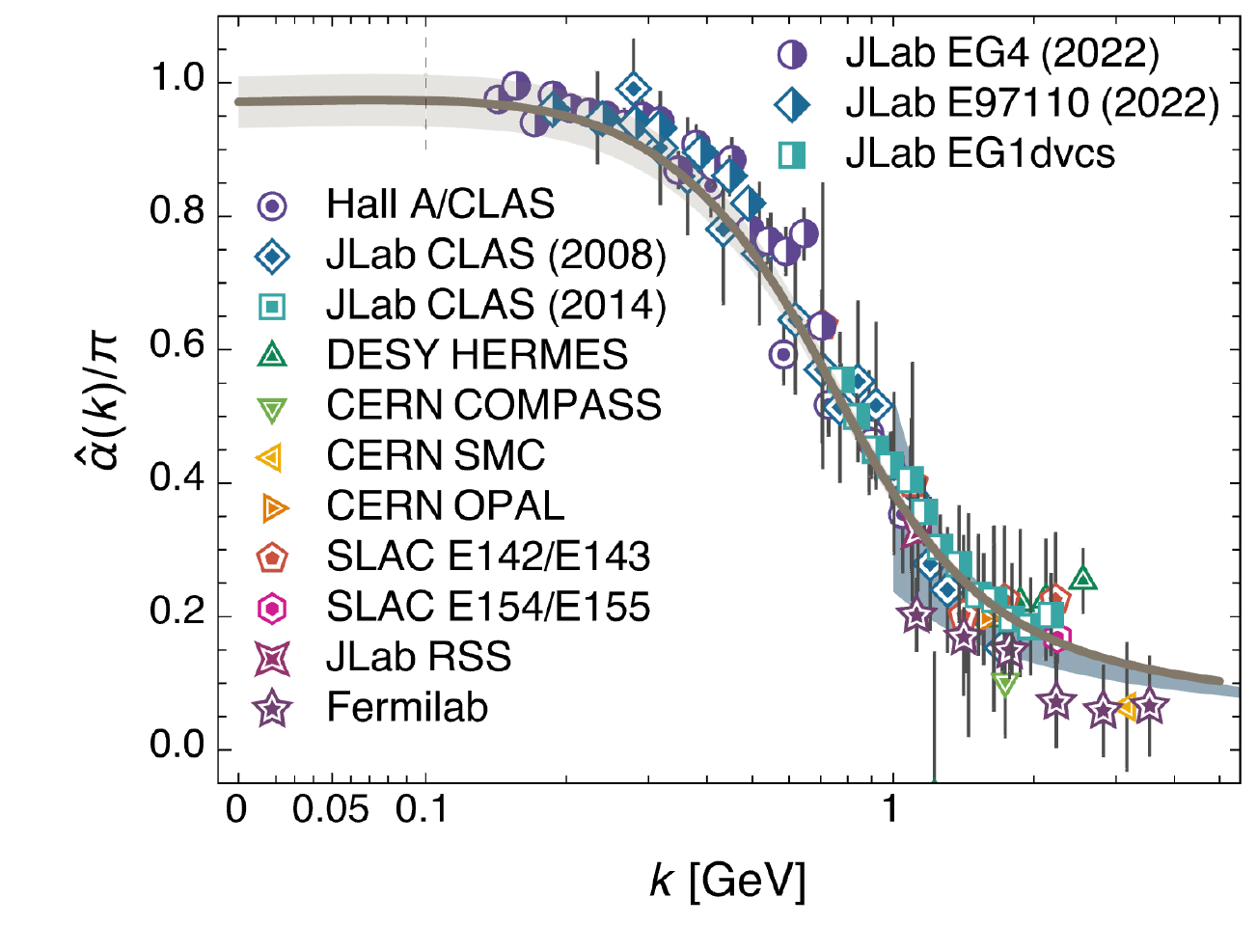}

\vspace*{-35ex}

\rightline{\parbox[t][10em][c]{0.44\textwidth}{
\caption{\label{Falpha}
$\hat{\alpha}(k)/\pi$ -- process-independent effective charge, calculated by combining results from continuum and lattice studies of QCD's gauge sector \cite{Cui:2019dwv}.
%
%
Existing data on the process-dependent charge $\alpha_{g_1}$, defined using the Bjorken sum rule, is shown for comparison.  Complete details and data sources are available in Refs.\,\cite{Deur:2022msf, Deur:2023dzc}.
(Image courtesy of D.\,Binosi.)}}}

\vspace*{5ex}

\end{figure}

During the past decade, however, the ghost-field hurdle has been overcome using a combination of pinch technique (PT) \cite{Cornwall:1981zr, Cornwall:1989gv, Pilaftsis:1996fh, Binosi:2009qm, Cornwall:2010upaFBS} and background field method (BFM) \cite{Abbott:1980hw, Abbott:1981ke}.  The PT+BFM approach facilitates a rearrangement and resummation of diagrams in a process that leads to a unique QCD running coupling, $\hat{\alpha}$, which is determined solely by a modified form of gluon vacuum polarisation \cite{Binosi:2016nme, Cui:2019dwv}, thereby delivering the desired QCD analogue of QED's Gell-Mann--Low coupling.  $\hat{\alpha}$ is RGI and PI.

The most up-to-date result for QCD's effective charge is drawn in Fig.\,\ref{Falpha}.  It was obtained \cite{Cui:2019dwv} by combining contemporary results from continuum analyses of QCD's gauge sector and lattice-QCD configurations generated with three domain-wall fermions at the physical pion mass \cite{Blum:2014tka, Boyle:2015exm, Boyle:2017jwu} to obtain a parameter-free prediction.

Notably,
\begin{equation}
\hat\alpha(k=0)/\pi=0.97(4)\,;
\end{equation}
so this coupling saturates to a large, finite value at $Q^2=0$.  The dynamical generation of a running gluon mass has eliminated the Landau pole that afflicts perturbative QCD.  With such behavior, QCD recovers the conformal character of the classical theory on $k\lesssim m_0/2$.  Thereupon, long-wavelength gluon modes have decoupled \cite{Brodsky:2008be}, so antiscreening ends.  This feature also resolves the problem of Gribov copies \cite{Gao:2017uox}, which might otherwise have precluded a nonperturbative definition of QCD.

The RGI+PI charge, $\hat\alpha(k)$, has been shown to deliver a unified description of many observables \cite{Roberts:2021nhw, Binosi:2022djx, Ding:2022ows}.  Significantly, too, $\hat\alpha(k)$ is pointwise practically identical to the effective charge $\alpha_{g_1}$ \cite{Deur:2022msf}, for which data is plentiful -- see Fig.\,\ref{Falpha}.  Thus, with $\hat\alpha(k)$, one has in hand an excellent candidate for that long-sought running coupling which characterises QCD interactions at all momentum scales \cite{Dokshitzer:1998nz}.  That being the case, then these results point strongly toward QCD being the first well-defined four-dimensional quantum field theory ever contemplated.

The running gluon mass in Fig.\,\ref{Fmassbudget}B leads to a massive gluon propagator.  Combined with $\hat\alpha(k)$, one then has the two principal elements required to form the kernel of the quark gap equation.  Solving that equation with the inputs described above, it is found that, without any tuning, light (even massless) quarks acquire a running mass, $M_{\hat m}(k)$, whose value at infrared momenta matches that which is usually identified with a constituent quark mass \cite[Sec.\,2C]{Roberts:2021nhw}.  The chiral-limit (no HB couplings into QCD, $\hat m=0$) result is displayed in Fig.\,\ref{Fmassbudget}B.  Its behaviour guarantees that quark+antiquark contributions also decouple from calculations of  $\hat\alpha(k\lesssim m_0/2)$, so screening also ends.

At this point, one has identified the three pillars that support the CSM paradigm of EHM: namely, the running gluon mass, process-independent effective charge, and running quark mass.  These quantities ensure and express the self-completing character of QCD.  Theory is today engaged in identifying entries in the huge array of their observable consequences -- a few of which will be sketched below -- and paths to measuring them.  The challenge for experiment is to test this body of predictions so that the boundaries of the Standard Model can finally be drawn.

\section{Charting EHM}
The proton was discovered one hundred years ago \cite{RutherfordI, RutherfordII, RutherfordIII, RutherfordIV}; and, being stable, it is the most ideal target in experiments.  Nevertheless, just as studying the ground state of the hydrogen atom did not reveal the need for and intricacies of QED, focusing on the ground state of only one form of hadron matter will not solve QCD.

A new era is approaching, with science poised to construct and begin operating high-luminosity, high-energy facilities that will enable detailed studies of new types and excited states of hadron matter.
For instance, one may anticipate a wealth of precise data that may finally reveal the inner workings of Nature's most fundamental Nambu-Goldstone bosons -- the $\pi$ and $K$ mesons \cite{Aguilar:2019teb, Chen:2020ijn, Anderle:2021wcy, Arrington:2021biu, Quintans:2022utc}.

Other opportunities can also be anticipated.  Baryons are the most fundamental three-body systems in Nature.  Amongst them, the proton is merely the simplest.  Quark models have been used to correlate properties of the tower of baryon states, but such models cannot explain them: they are too simple and break too many symmetries -- e.g., see Sec.\,\ref{SecEB}.  QCD is a Poincar\'e-invariant quantum non-Abelian gauge theory; and theory must be developed to a point from which it can explain how QCD itself builds each of the baryons in the complete spectrum.  As a beginning, it must elucidate the impacts of running-couplings and -masses on baryon properties and interactions.

\subsection{Faddeev equation}
\label{SecFE}
\begin{figure}[t]
\vspace*{3ex}

\begin{tabular}{ll}
\includegraphics[clip, width=0.55\textwidth]{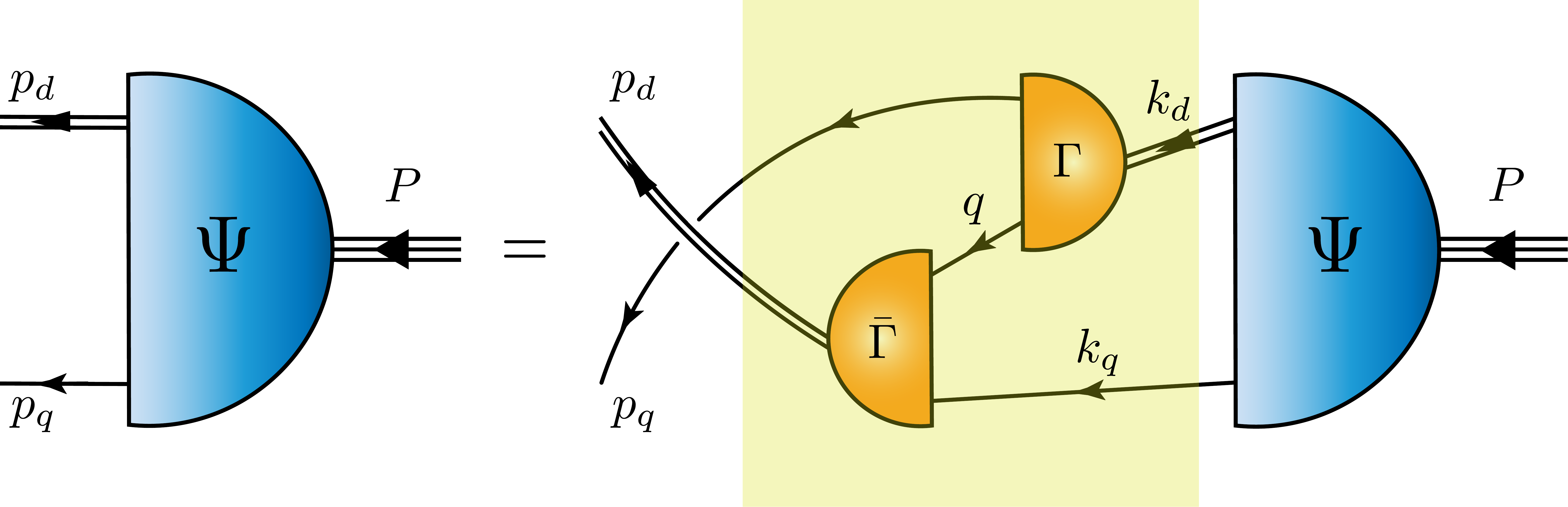}& \\[-23ex]
& \includegraphics[clip, width=0.4\textwidth]{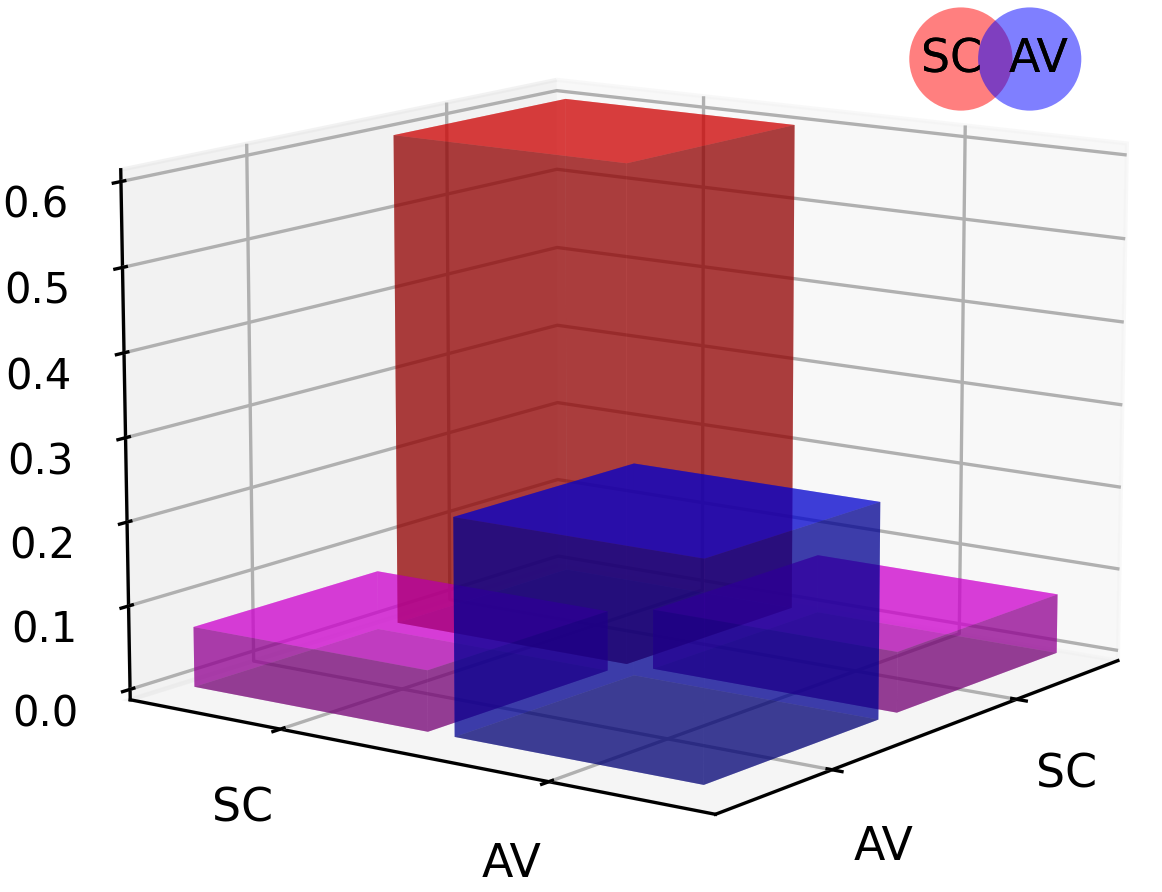} \\[-4ex]
(\textbf{A}) & (\textbf{B})
\end{tabular}
\vspace*{1ex}

\caption{\label{figFaddeev}
\textbf{A}. Quark+diquark Faddeev equation.  The solution, $\Psi$, is the Poincar\'e-covariant, matrix-valued Faddeev amplitude for a baryon with total momentum $P=p_q+p_d=k_q+k_d$ constituted from three valence quarks, two of which are always contained in a nonpointlike diquark correlation.  $\Psi$ expresses the relative momentum correlation between the dressed-quarks and -diquarks.
Legend. \emph{Shaded rectangle} -- Faddeev kernel; \emph{single line} -- dressed-quark propagator; $\Gamma$ -- diquark correlation amplitude; and \emph{double line} -- diquark propagator.
Regarding ground-state nucleons ($n$ - neutron, $p$ - proton), both contain isoscalar-scalar diquarks, $[ud]\in(n,p)$, and isovector-axialvector diquarks $\{dd\}\in n$, $\{ud\}\in (n,p)$, $\{uu\}\in p$.
\textbf{B}.  Breakdown of the proton charge into contributions from various diquark correlations.  Whilst the $[ud]$ (SC) is dominant, there is a material contribution ($\approx 40$\%) from $\{uu\}$ and $\{ud\}$ correlations (AV): both direct AV$\times$AV and via constructive SC$\times$AV interference.
}
\end{figure}

A Poincar\'e-covariant Faddeev equation for baryons was introduced thirty-five years ago \cite{Cahill:1988dx, Reinhardt:1989rw, Efimov:1990uz}.  As a problem describing three dressed valence quarks bound by exchange of dressed gluons, it was first solved in Ref.\,\cite{Eichmann:2009qa}.  In that guise, today, it has been used to deliver a range of results on baryon spectra and interactions \cite{Eichmann:2016yit, Sanchis-Alepuz:2017mir, Qin:2018dqp, Wang:2018kto, Qin:2019hgk}.  Notwithstanding steady progress along that path, many Poincar\'e-covariant studies of baryon properties continue to profit from an important corollary of EHM; namely, any quark+antiquark interaction that produces a good description of meson properties also generates strong, nonpointlike quark+quark (diquark) correlations in multiquark systems \cite{Cahill:1987qr}.  A contemporary discussion of the origin, impact of and empirical evidence for such diquark correlations is presented elsewhere \cite{Barabanov:2020jvn}.   Capitalising on the emergence of diquarks, a fully-interacting quark+nonpointlike-diquark approximation to the Faddeev equation -- sketched in Fig.\,\ref{figFaddeev}A -- is used for most applications.

One of the key predictions made by solutions of the Faddeev equation in Fig.\,\ref{figFaddeev}A is that ground-state nucleons contain \emph{both} $(I,J^P) = (0,0^+)$ and $(1,1^+)$ diquark correlations.  In fact, as illustrated in Fig.\,\ref{figFaddeev}B, the isoscalar-axialvector diquark is responsible for roughly 40\% of the proton's unit charge.

\subsection{Nucleon axial form factor}
\label{SecAxial}
The Faddeev equation in Fig.\,\ref{figFaddeev}A has provided the foundation for calculations of baryon elastic and transition electromagnetic form factors since the development of an associated symmetry-preserving electromagnetic current more than twenty years ago \cite{Oettel:1999gc}.  However, an analogous axialvector current has only recently become available \cite{Chen:2020wuq, Chen:2021guo}.  In light of widespread international efforts focused on neutrino physics, this represents crucial progress because CSMs can now be used to deliver predictions for the nucleon axial form factor.  $G_A(Q^2)$ is a key element in the analysis and reliable interpretation of neutrino experiments, which depend upon sound theoretical knowledge of neutrino/antineutrino-nucleus ($\nu/\bar \nu$-$A$) interactions \cite{Mosel:2016cwa, Alvarez-Ruso:2017oui, Hill:2017wgb, Lovato:2020kba, King:2020wmp} and such, in turn, rely heavily upon $G_A(Q^2)$.

\begin{figure}[t]

\begin{tabular}{ll}
\includegraphics[clip, width=0.475\textwidth]{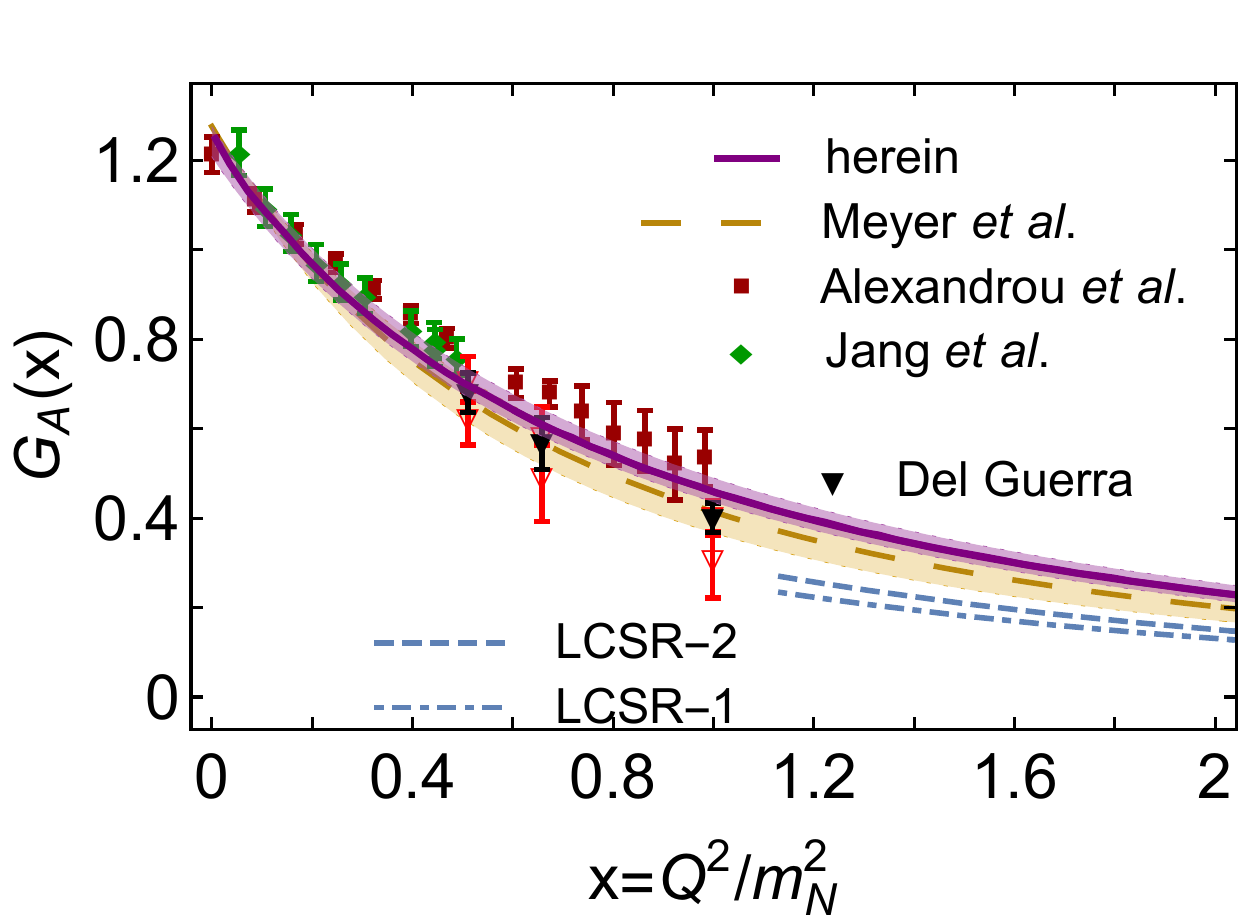}&
\includegraphics[clip, width=0.475\textwidth]{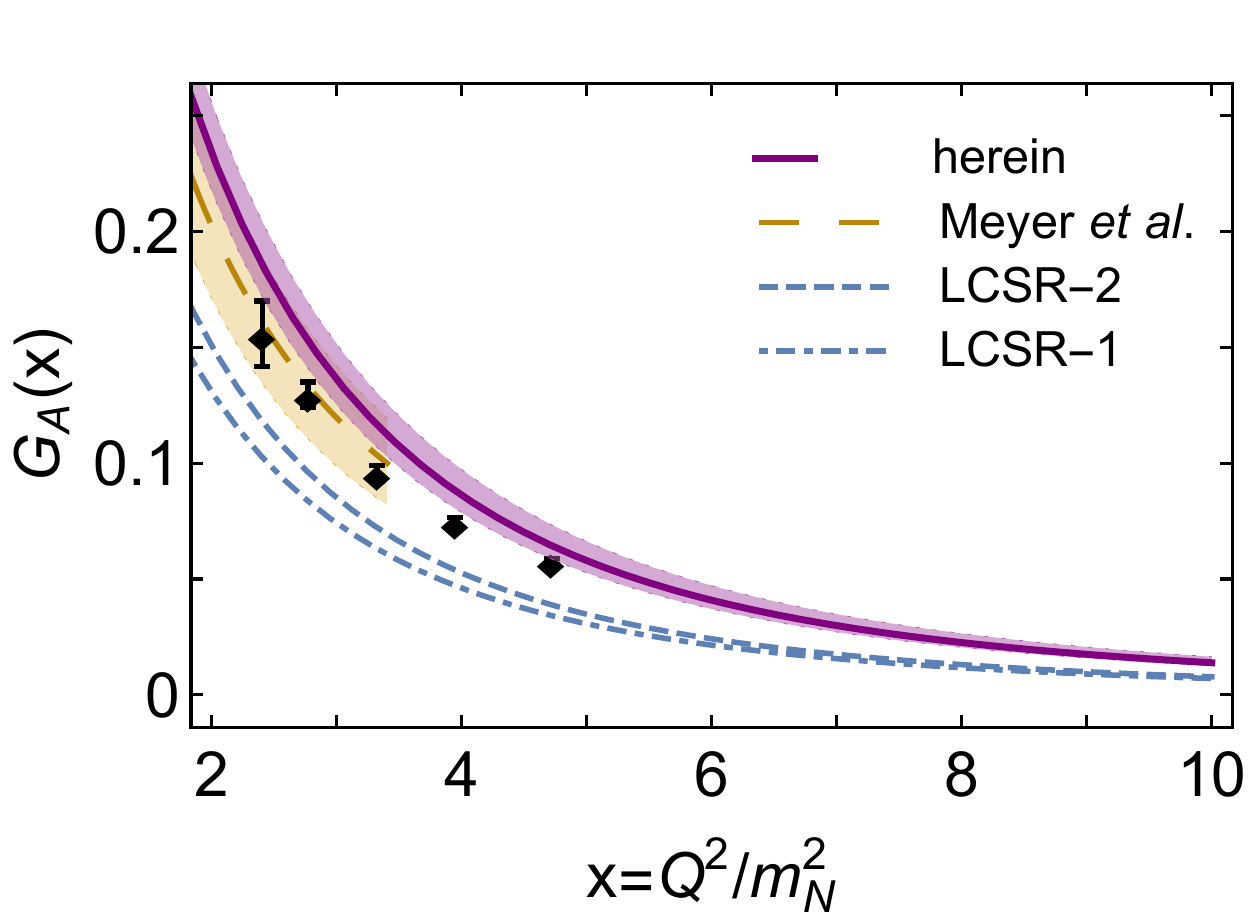} \\[-4ex]
(\textbf{A}) & (\textbf{B})
\end{tabular}
\vspace*{1ex}

\caption{\label{FigGAx}
\textbf{A}.
Low-$x$ behaviour of $G_A(x=Q^2/m_N^2)$.
Solid purple curve -- CSM prediction  \cite{Chen:2020wuq, Chen:2021guo}.  The bracketing shaded band shows the sensitivity to $\pm 5$\% variation in the masses of the scalar and axialvector diquarks:
$m_{[ud]} = 0.80\,\mbox{GeV}$, $m_{\{uu\}} = m_{\{ud\}} =0.89\,$GeV.
Comparisons provided with:
dipole fit to data -- long-dashed gold curve and like-coloured band \cite{Meyer:2016oeg};
values deduced from threshold pion electroproduction data under three different assumptions \cite{DelGuerra:1976uj} -- filled and open triangles;
results from lattice QCD --
green diamonds \cite{Jang:2019vkm} and
red squares \cite{Alexandrou:2017hac};
and light-cone sum rule (LCSR) results \cite{Anikin:2016teg} obtained using two forms of the nucleon distribution amplitudes \cite[Table~I]{Anikin:2013aka}. -- LCSR-1 (dot-dashed blue curve) and LCSR-2 (dashed blue curve).  LCSRs cannot deliver a result below $x \simeq 1$.
\textbf{B}.
Large-$x$ behaviour of $G_A(x)$ \cite{ChenChen:2022qpy}.
Curves drawn consistent with the Panel A legend and additionally include threshold pion electroproduction data from Ref.\,\cite{CLAS:2012ich} -- black diamonds.
}
\end{figure}

The recent CSM prediction for $G_A(x=Q^2/m_N^2)$ is drawn in Fig.\,\ref{FigGAx}.  Compared with the lattice-QCD result in Ref.\,\cite{Jang:2019vkm} on their domain of overlap, the mean-$\chi^2=0.27$; namely, there is excellent agreement.  On the other hand, as highlighted by Fig.\,\ref{FigGAx}, whereas lattice calculations are only available on $0 \lesssim Q^2  \lesssim m_N^2$, the reach of CSM predictions is far greater, with results now available to $Q^2 = 10\, m_N^2$.  In connection with Fig.\,\ref{figFaddeev}B, it is worth observing that, insofar as $g_A=G_A(x=0)$ is concerned, $71$\% of its value is produced by the weak-boson striking the dressed-quark in association with a spectator scalar diquark.  The remaining 29\% owes to the nucleon's axialvector diquark component, with almost half of this amount produced by weak-interaction-induced SC$\leftrightarrow$AV transitions.

At the structural level, the isovector quantity $g_A$ measures the difference between the light-front number-density of quarks with helicity parallel to that of the nucleon and the density of quarks with helicity antiparallel \cite{Chang:2012cc}.  This makes it interesting to provide a flavour separation of the proton axial charge.  In the absence of axialvector diquark correlations in the nucleon, $g_A^d/g_A^u = -0.054(13)$.  On the other hand, using the complete nucleon amplitude, with the diquark content illustrated in Fig.\,\ref{figFaddeev}B, $g_A^d/g_A^u = -0.32(2)$.  What does experiment have to say?  If one assumes SU$(3)$-flavour symmetry, then the axial charges of octet baryons can be expressed in terms of two low-energy constants,  $D$, $F$; and $g_A^u = 2 F$, $g_A^d=F-D$.  Using available empirical information \cite{Workman:2022ynf}, one finds $D= 0.774(26)$, $F=0.503(27)$; so, $g_A^d/g_A^u= -0.27(4)$.  Comparing this value with the CSM prediction, one may conclude that the likelihood of a scalar diquark only picture of the proton being consistent with data is just $1/7,100,000$.

The flavour-singlet combination of flavour-separated contributions to the proton axial form factor gives the fraction of the proton spin carried by valence quarks.  This quantity is weakly dependent on the resolving scale, $\zeta$.  Referred to the hadron scale, $\zeta=\zeta_{\cal H}$, whereat all properties of the system are carried by dressed valence degrees-of-freedom, $g_A^u+g_A^d = 0.65(2)$ \cite{ChenChen:2022qpy}; namely, at $\zeta_{\cal H}$, quarks carry roughly 65\% of the proton spin.  The remainder is lodged with quark+diquark orbital angular momentum.  Consistent with this picture, an extension of the analysis to the entire octet of ground-state baryons using a symmetry-preserving treatment of a contact interaction indicates \cite{Cheng:2022jxe} that dressed quarks carry 50(7)\% of the spin of these systems at $\zeta_{\cal H}$.

Shifting focus to Fig.\,\ref{FigGAx}B, it becomes apparent that only CSMs are able to deliver a prediction for $G_A(x)$ on the entire domain of spacelike momenta: LCSRs cannot reliably be employed on $Q^2\lesssim m_N^2$ and algorithmic challenges currently preclude the use of lattice-QCD on $Q^2\gtrsim m_N^2$.
It is important, therefore, that the parameter-free CSM prediction agrees with all available data, both at low and high $Q^2$.
The highest $Q^2$ data set was obtained using large momentum transfer threshold pion electroproduction \cite{CLAS:2012ich}.
This experimental technique could potentially be used to reach even higher $Q^2$ values; so, test the CSM picture of proton short-distance axial structure.
It is also worth noting that the oft-used dipole fitting \emph{Ansatz} provides a fair representation of $G_A(x)$ data on $x\in [0,3]$.  However, outside the fitted domain, the quality of the approximation deteriorates rapidly: at $x=10$, the dipole overestimates the true result by 56\%.

\begin{figure*}[!t]
\begin{tabular}{lcl}
%
%
\includegraphics[clip, width=0.405\textwidth]{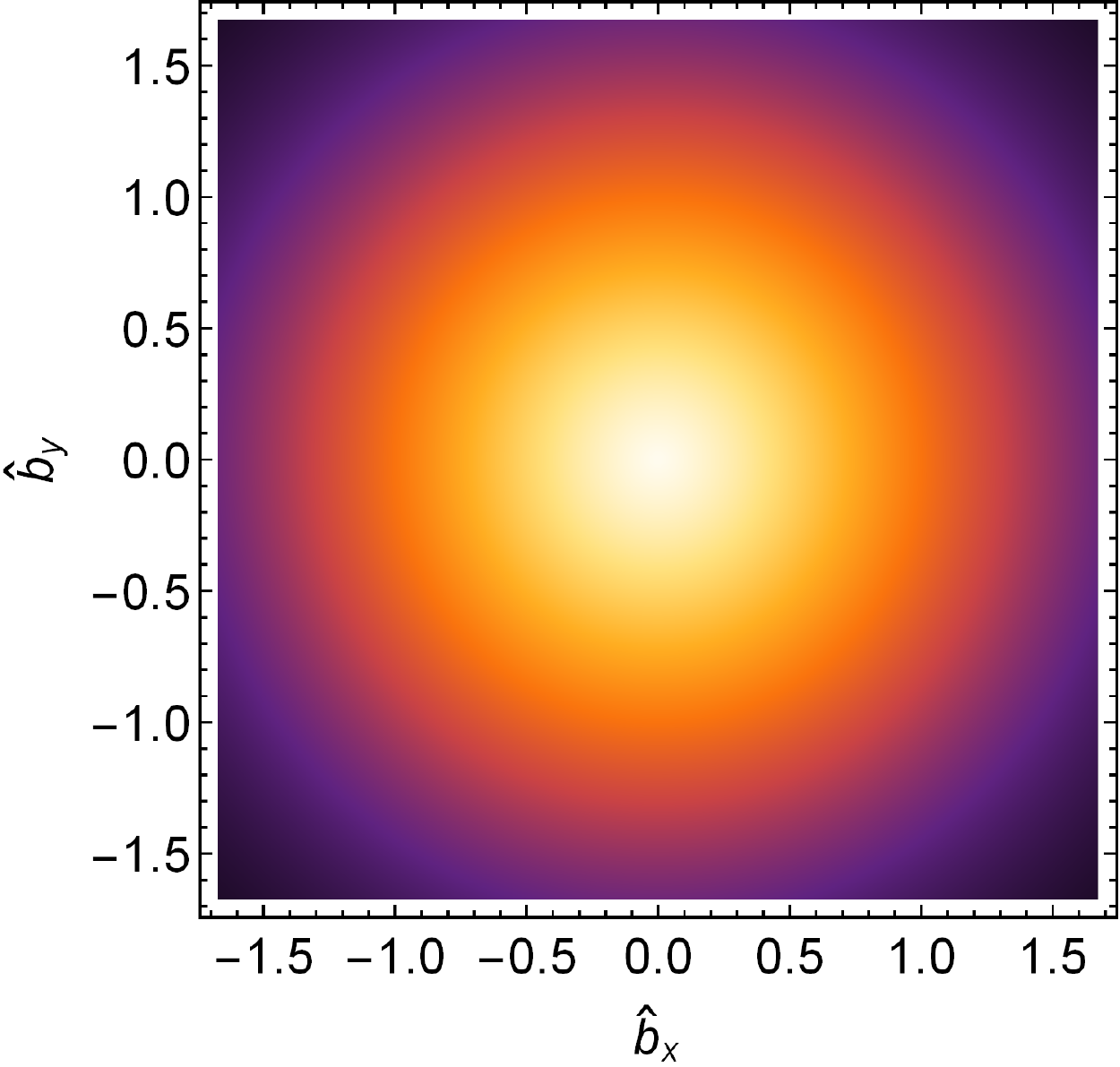} &
\hspace*{1em} \includegraphics[clip, width=0.06\textwidth]{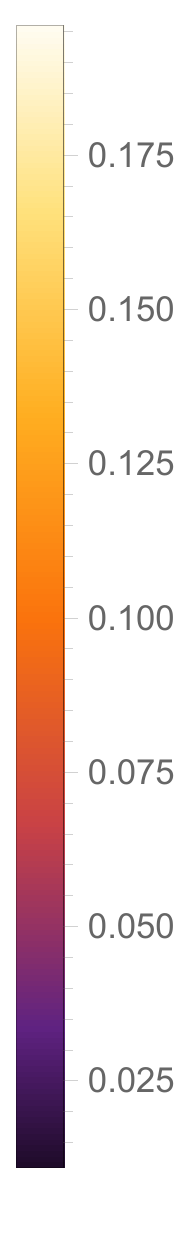} \hspace*{0em} &
\includegraphics[clip, width=0.405\textwidth]{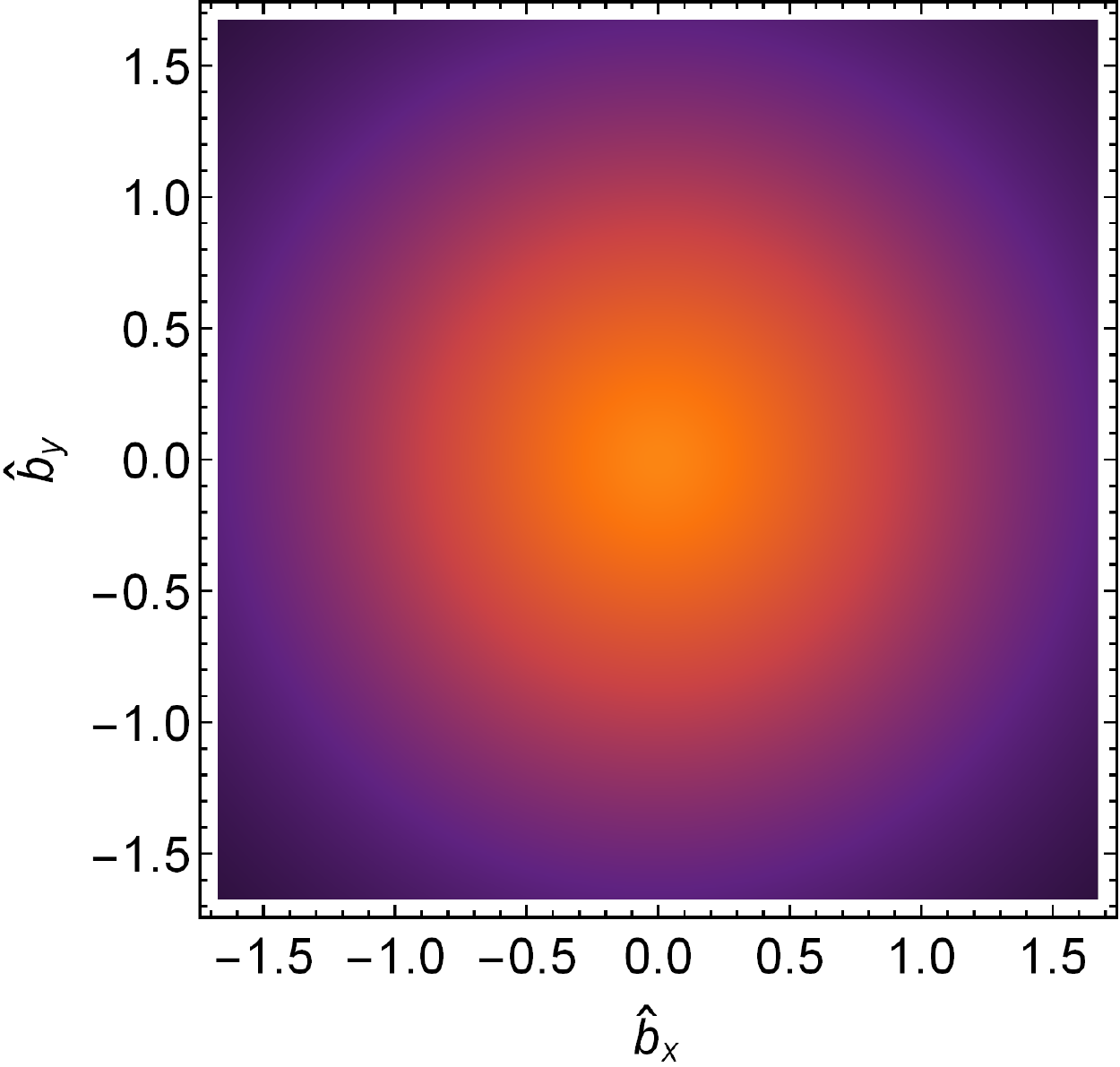}\\[-1ex]
{(\textbf{A}) \hspace*{1em}$0^+\!, d $} & & {(\textbf{B}) \hspace*{1em}$0^+\!, u $}
\end{tabular}
\vspace*{1ex}

\begin{tabular}{lcl}
\includegraphics[clip, width=0.405\textwidth]{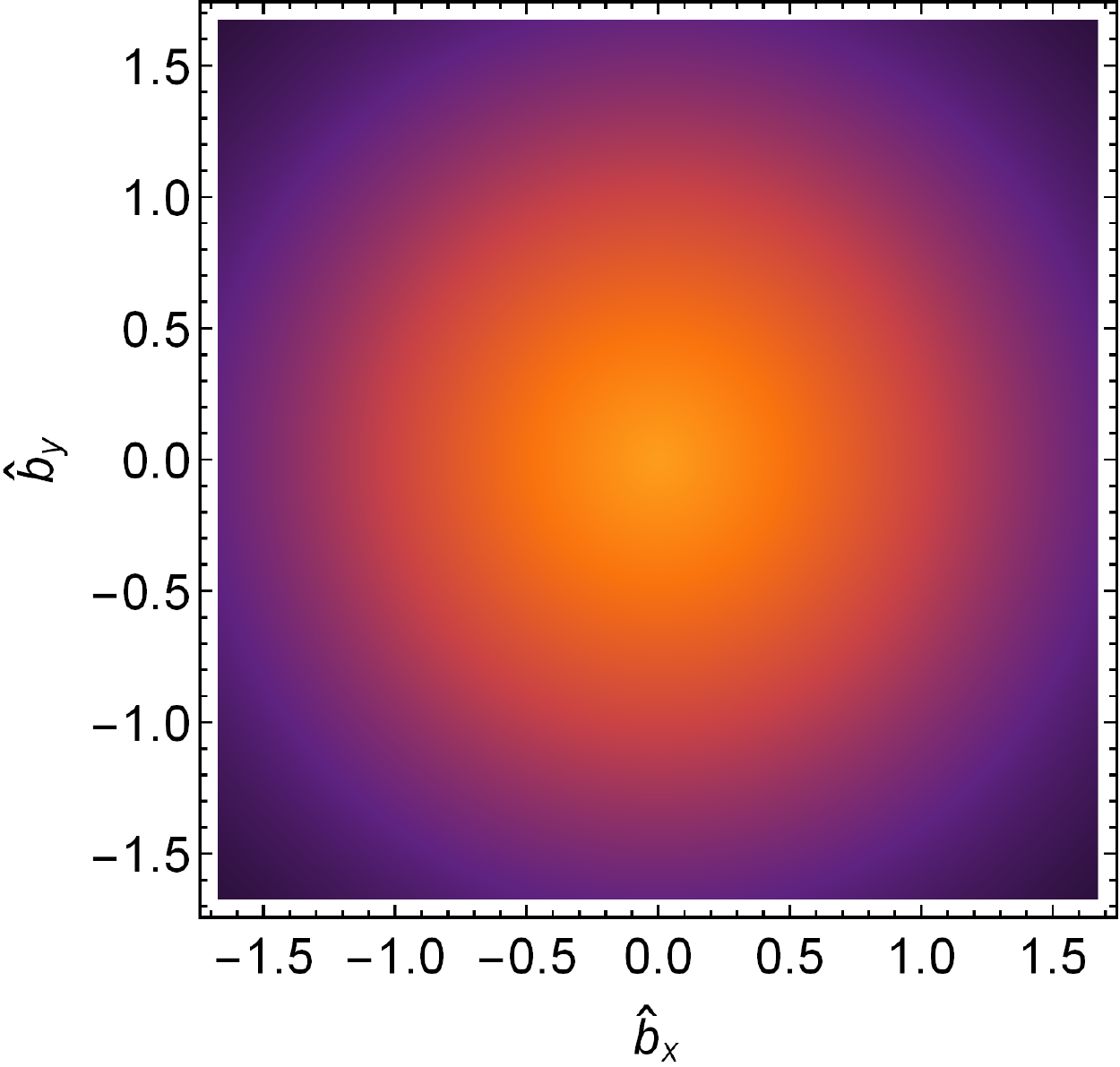} &
\hspace*{1em} \includegraphics[clip, width=0.054\textwidth]{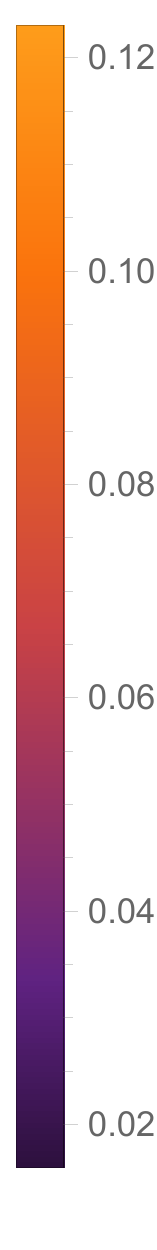} \hspace*{0.4em} &
\includegraphics[clip, width=0.405\textwidth]{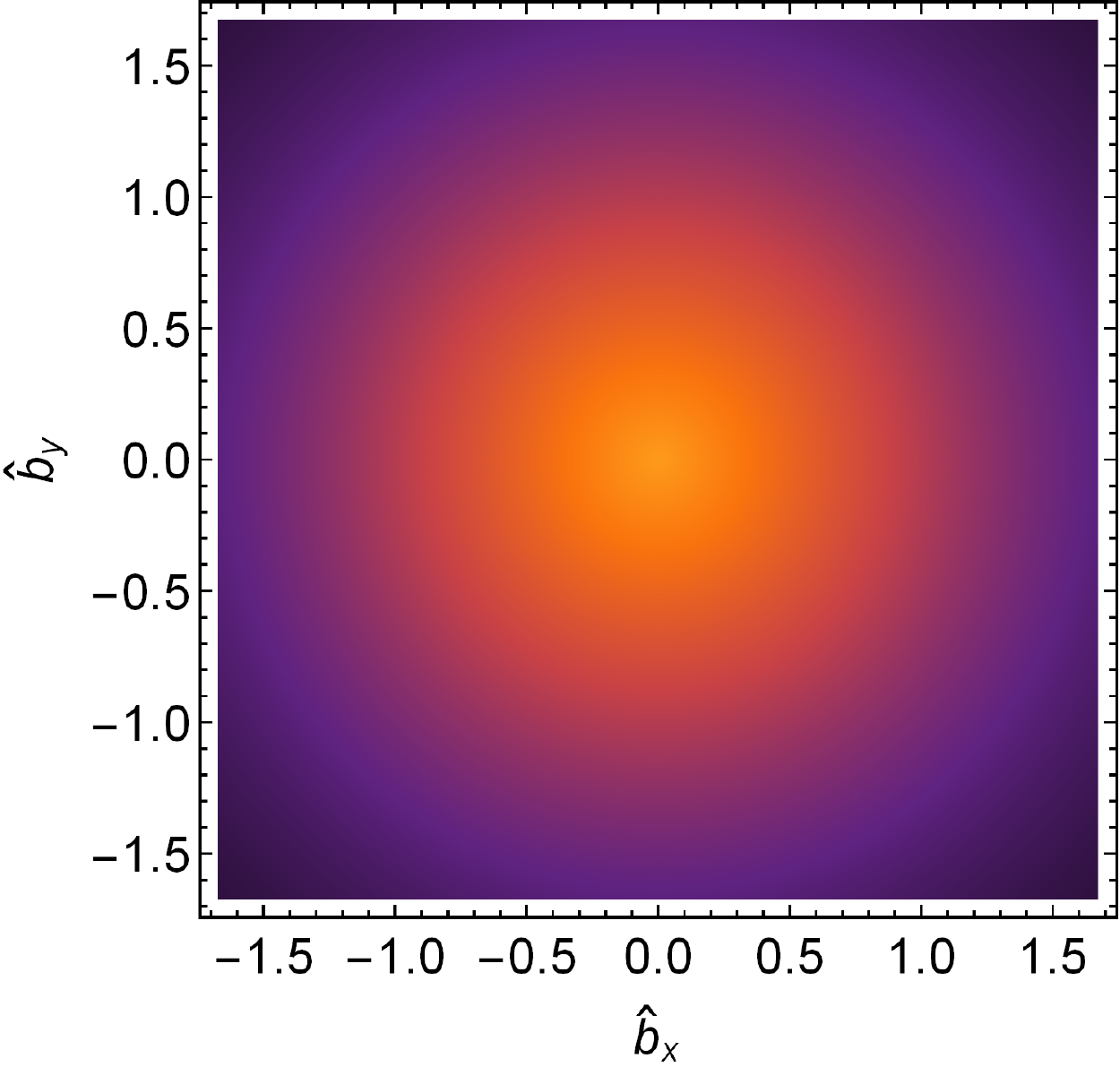}\\[-1ex]
{(\textbf{C}) \hspace*{1em}$0^+$\&$1^+\!, d $} & & {(\textbf{D}) \hspace*{1em}$0^+$\&$1^+\!, u $}
\end{tabular}
\vspace*{1ex}

\caption{\label{figdensity2D}
Transverse density profiles,  Eq.\,\eqref{density}: $\hat\rho_A^f(|\hat b|)/g_A^f$, $f=d,u$.
Panels A and B calculated from the flavour separated proton axial form factors for a scalar-diquark-only proton, whereas
Panels C and D employ the complete proton wave function, including axialvector diquarks.
Removing the $1/g_A^f$ normalisation, the $b=0$ values are: (A) $\hat \rho_A^d(0) = -0.009$; (B) $\rho_A^u(0)= 0.097$; (C) $\hat \rho_A^d(0) = -0.038$; (D) $\rho_A^u(0)= 0.12$.
\emph{N.B}.\ $\int d^2 \hat b \,\hat\rho_A^f(|\hat b|)/g_A^f = 1$, $f=u,d$.
}
\end{figure*}

With flavour-separated results for the nucleon axial form factor to large $Q^2$, it is possible to calculate proton light-front transverse axial charge spatial density profiles:
\begin{equation}
\label{density}
\hat\rho_A^f(|\hat b|) = \int\frac{d^2 \vec{q}_\perp}{(2\pi)^2}\,{\rm e}^{i \vec{q}_\perp \cdot\hat b}G_A^f(x)\,,
\end{equation}
with $G_A^f(x)$ interpreted in a frame defined by $Q^2 = m_N^2 q_\perp^2$, $m_N q_\perp = (\vec{q}_{\perp 1},\vec{q}_{\perp 2},0,0)=(Q_1,Q_2,0,0)$.  Here, $|\hat b|$ and $\hat\rho_A^f $ are dimensionless.  The results \cite{ChenChen:2022qpy} are displayed in Fig.\,\ref{figdensity2D}.
Omitting axialvector diquarks -- Panels A and B -- the magnitude of the $d$ quark contribution to $G_A(x)$ is just 10\% of that from the $u$ quark.  Furthermore, the $d$ quark distribution is far more localised: comparing transverse radii, $r_{A_d}^\perp \approx 0.5 r_{A_u}^\perp$.
Working instead with the realistic Faddeev amplitude associated with Fig.\,\ref{figFaddeev}B  -- Panels C and D, then the $d$ quark profile is only somewhat more pointlike than that of the $u$ quark: $r_{A_d}^\perp \approx 0.9 r_{A_u}^\perp$.  In addition, the magnitude of the $d$ quark contribution to $G_A(x)$ is roughly 30\% of that from the $u$ quark.
These results further highlight the key role played by axialvector diquarks in nucleon structure.  They cannot realistically be neglected.

\subsection{Excited baryons}
\label{SecEB}
As stressed above, QCD is a Poincar\'e-invariant quantum non-Abelian gauge field theory.
It follows that the wave function of each and every QCD bound-state must be Poincar\'e covariant.
So, irrespective of the quark model assignments given, e.g., to hadrons in Ref.\,\cite{Workman:2022ynf}, every hadron contains orbital angular momentum.
For the proton, this was made clear above in connection with the flavour separation of its axial charge.
The same is even true of the $J^P=0^-$ pion, whose rest-frame wave function contains two \textsf{S}-wave and two \textsf{P}-wave components \cite{Maris:1997tm, Bhagwat:2006xi}.
Moreover, no system is simply the radial excitation of another.
On top of these things, there is no simple connection in quantum field theory between parity and orbital angular momentum.
Parity is a Poincar\'e invariant quantum number.
$L$ is not Poincar\'e invariant.  Its value depends on the observer's frame of reference, i.e., every separation of $J$ into $L+S$ is frame dependent.
Hence, e.g., negative parity excited states cannot simply be orbital angular momentum excitations of positive parity ground states.

It should now be plain that the QCD structure of hadrons -- mesons and baryons -- is far richer than can be produced by quark models, relativised or not.  This elevates the problem of describing hadron structure within QCD to the highest level.  Aspects of the meson problem are discussed elsewhere \cite{Xu:2022kng}.  Baryons are even more of a challenge.  They are the most fundamental three-body systems in Nature; and if we don't understand how QCD, a Poincar\'e-invariant quantum field theory, builds each of the baryons in the complete spectrum, then we don't understand Nature.

\begin{figure*}[!t]
\hspace*{-1ex}\begin{tabular}{lcl}
%
\includegraphics[clip, width=0.44\textwidth]{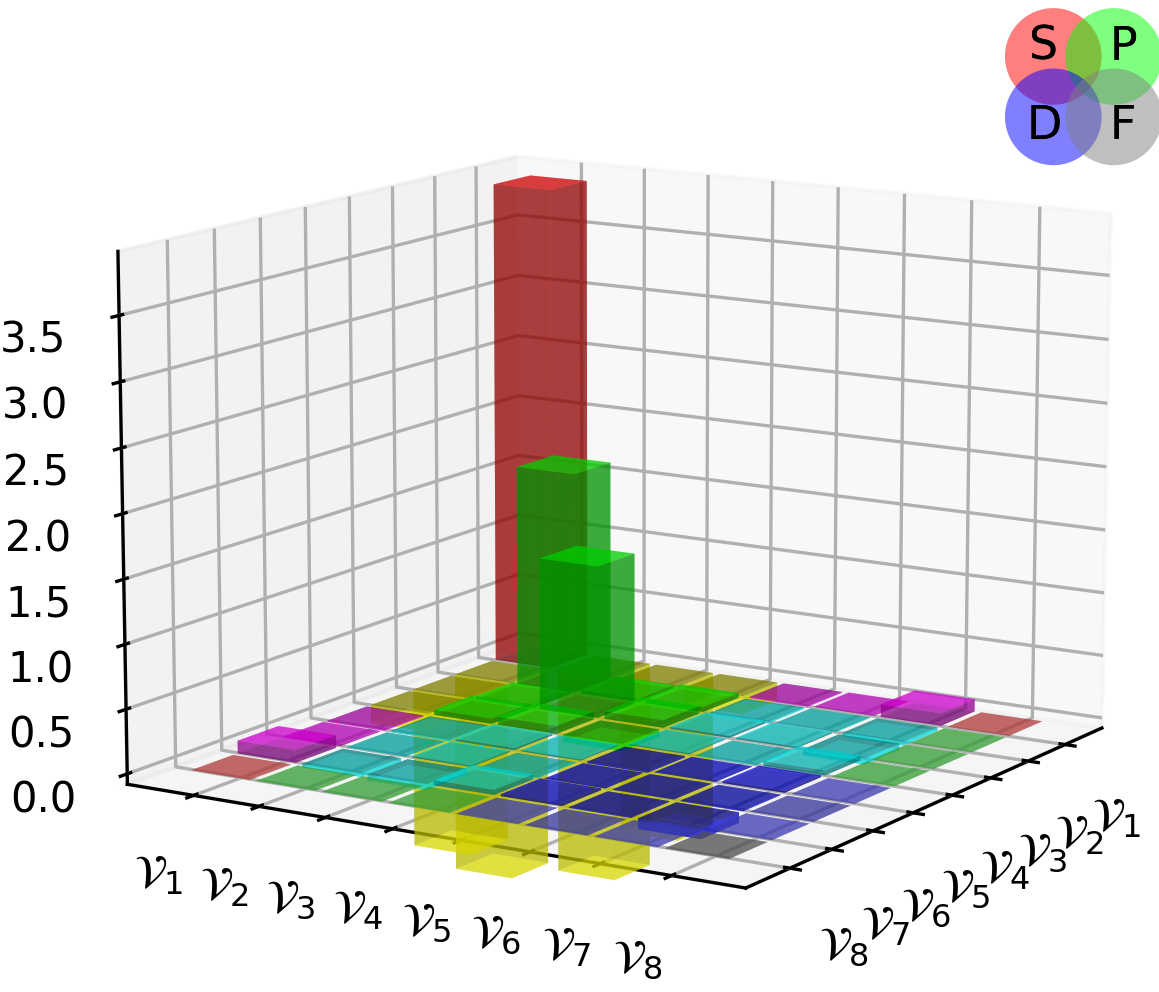} & \hspace*{2em} &
\includegraphics[clip, width=0.44\textwidth]{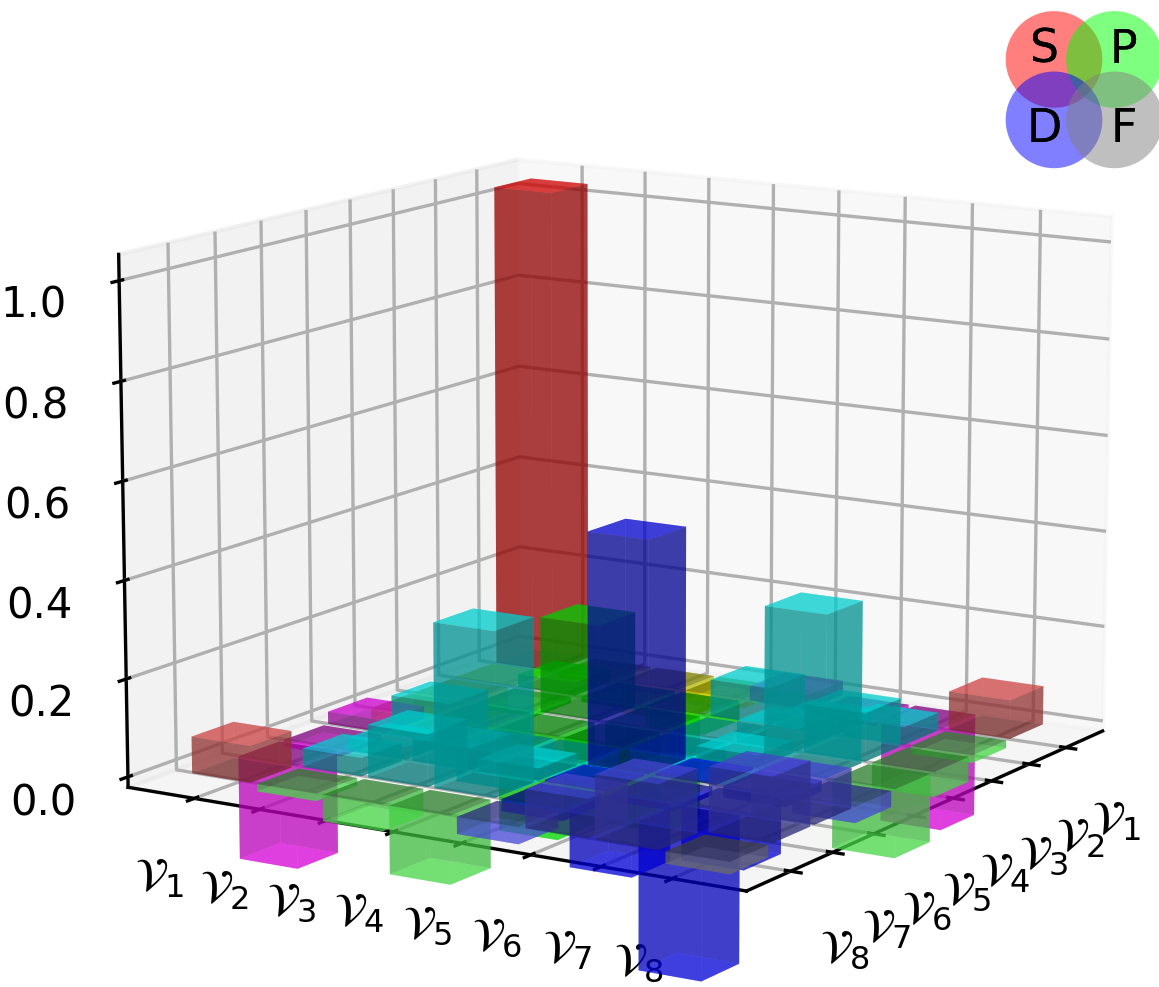} \\[-1ex]
{(\textbf{A})} & & {(\textbf{B})}\\
\includegraphics[clip, width=0.44\textwidth]{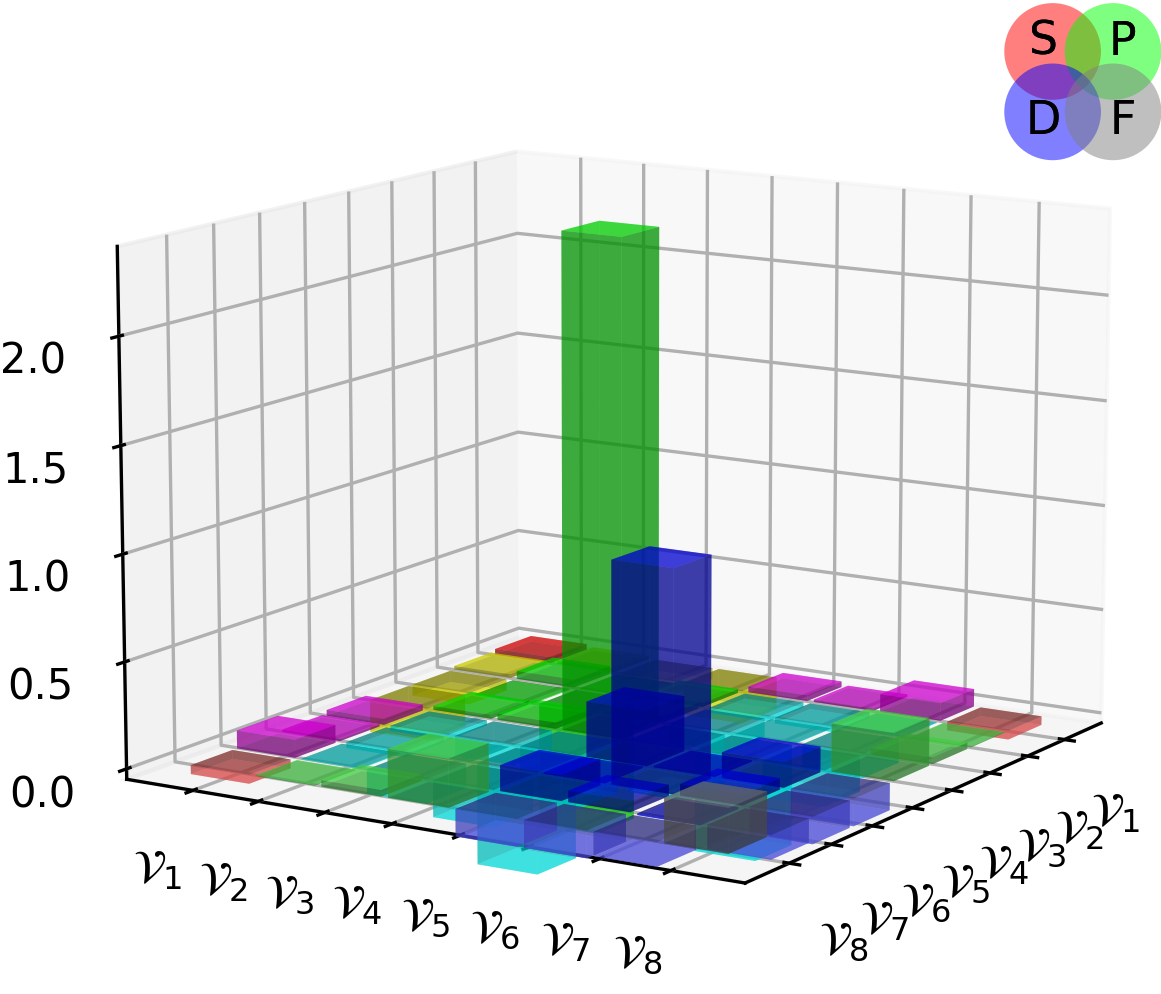} & \hspace*{2em} &
\includegraphics[clip, width=0.44\textwidth]{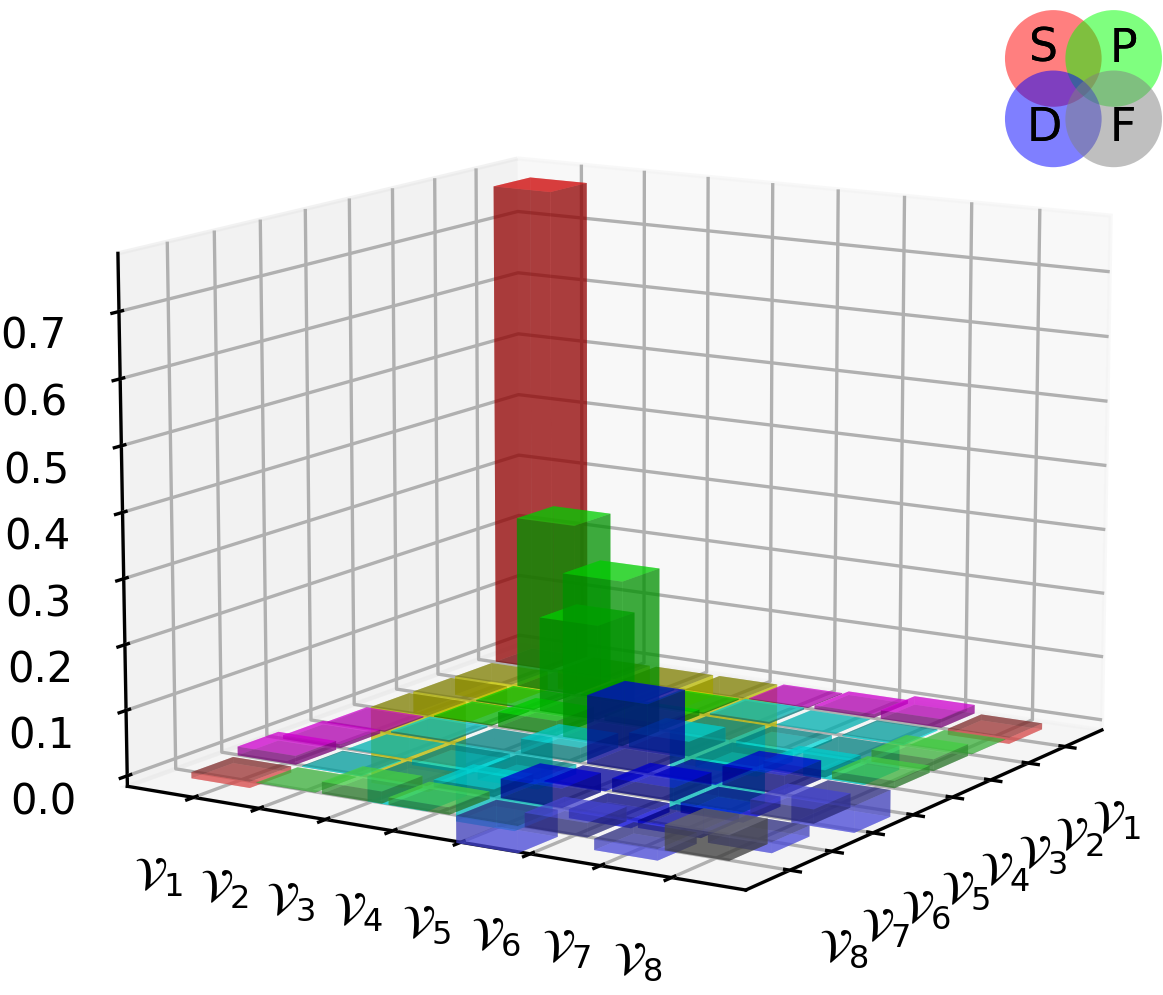} \\
{(\textbf{C})} & & {(\textbf{D})}
\end{tabular}
\caption{\label{LFigures}
Rest frame quark+$(1,1^+)$-diquark orbital angular momentum content of $(\tfrac{3}{2},\tfrac{3}{2}^\pm)$ states considered in Ref.\,\cite{Liu:2022ndb}, as measured by the contribution of the various components to the associated wave function canonical normalisation constant:
{\bf A} -- $ \Delta(1232)\tfrac{3}{2}^+$;
{\bf B} -- $ \Delta(1600)\tfrac{3}{2}^+$;
{\bf C} -- $ \Delta(1700)\tfrac{3}{2}^-$; and
{\bf D} -- $ \Delta(1940)\tfrac{3}{2}^-$.
There are both positive (above plane) and negative (below plane) contributions to the normalisations, each of which is positive overall.
(Interpretative legend drawn in Fig.\,\ref{LWFlegend}.)}
\end{figure*}

These points are highlighted in two recent Faddeev equation studies of baryons \cite{Liu:2022ndb, Liu:2022nku}.  Owing to the simpler character of $\Delta$-like baryons, the former will be used as the illustrator herein.  For instance, although $(\tfrac{3}{2},\tfrac{3}{2}^\pm)$ states can contain $(1,1^+)$ and $(1,1^-)$ diquark correlations, one may neglect the latter and still arrive at a reliable description.

\begin{figure}[h]

\includegraphics[clip, width=0.52\textwidth]{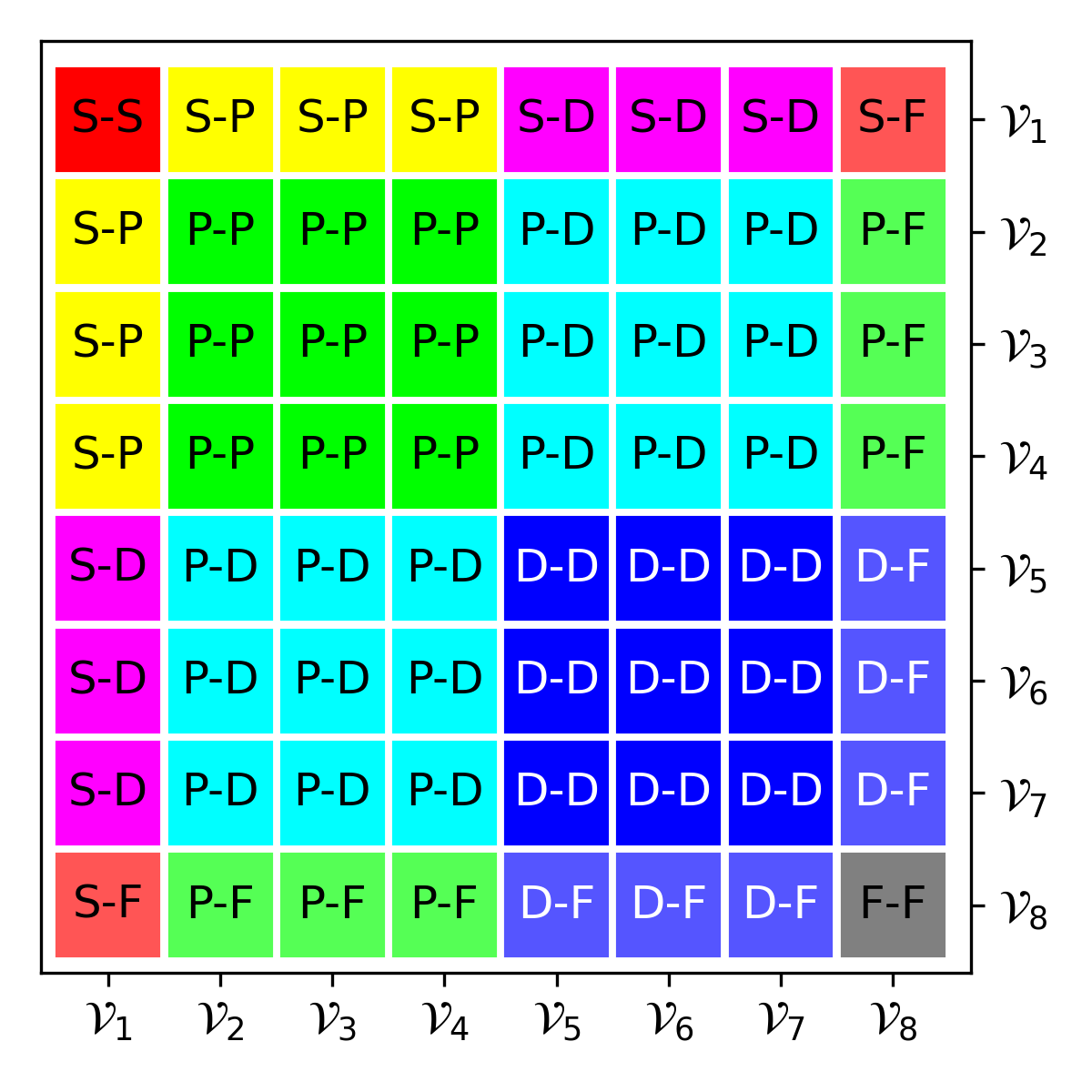}

\vspace*{-38ex}

\rightline{\parbox[t][10em][c]{0.44\textwidth}{
\caption{\label{LWFlegend}
Legend for interpretation of Figs.\,\ref{LFigures}A\,--\,D, identifying interference be\-tween the various identified orbital angular momentum basis components in the baryon rest frame.}}}

\vspace*{8ex}

\end{figure}

Considering first the $(\tfrac{3}{2},\tfrac{3}{2}^+)$ states, rest-frame quark+diquark angular momentum decompositions of their wave function canonical normalisation constants are drawn in the top row of Fig.\,\ref{LFigures}.
Regarding the $\Delta(1232) \tfrac{3}{2}^+$ -- Fig.\,\ref{LFigures}A, the normalisation is primarily determined by $\mathsf S$-wave components, but there are significant, constructive $\mathsf P$ wave contributions and also strong $\mathsf S\otimes \mathsf P$-wave destructive interference terms.  This structural picture is confirmed by comparisons with data on the $\gamma+p \to \Delta(1232)$ transition form factors \cite{Eichmann:2011aa, Segovia:2014aza, Lu:2019bjs}.

Regarding the $\Delta(1600)\tfrac{3}{2}^+$ -- Fig.\,\ref{LFigures}B, $\mathsf S$-wave contributions to the normalisation are dominant, but there are also prominent $\mathsf D$-wave components, material $\mathsf P \otimes \mathsf D$-wave interference contributions, and numerous $\mathsf F$-wave induced interference terms.
Strong higher partial waves are also seen in related three-body Faddeev equation studies of this baryon \cite{Eichmann:2016hgl, Qin:2018dqp}.
Looking closely at the $\Delta(1600)\tfrac{3}{2}^+$ wave function \cite{Liu:2022ndb}, the state has the appearance of the first radial excitation of the $\Delta(1232) \tfrac{3}{2}^+$; but, evidently, this excitation doesn't have the simplicity of a quark model state.
The quark+diquark structural picture of the $\Delta(1600)\tfrac{3}{2}^+$ has been used to calculate $\gamma+p \to \Delta(1600)$ transition form factors \cite{Lu:2019bjs}.  Those predictions are consistent with preliminary analyses of $\pi^+ \pi^- p$ electroproduction data collected at Jefferson Lab \cite[Fig.\,9]{Carman:2023zke}.

The $\Delta(1700)\tfrac{3}{2}^-$ normalisation strengths are shown in Fig.\,\ref{LFigures}C: $\mathsf P$-wave components are dominant, but $\mathsf D$-wave and $\mathsf P \otimes \mathsf D$ interference is evident, and also some $\mathsf D \otimes \mathsf F$ contributions.  $\Delta(1700)\tfrac{3}{2}^-$ electrocoupling data are available from Jefferson Lab \cite{Burkert:2002zz, CLAS:2009tyz, Mokeev:2013kka}.  However, their current reach in $Q^2$ is insufficient to test these $\Delta(1700)\tfrac{3}{2}^-$ structure predictions.

The biggest surprise is provided by the $\Delta(1940)\tfrac{3}{2}^-$, whose normalisation strengths are displayed in Fig.\,\ref{LFigures}D.
The image reveals that this state is possibly a peculiar system, \emph{viz}.\ a negative-parity baryon whose rest-frame wave function is largely $\mathsf S$-wave in character.
Whilst such an outcome is impossible in quantum mechanics quark models, it is readily achievable in Poincar\'e-invariant quantum field theory owing to the connection between the Dirac matrix $\gamma_5$ and parity.
Some may nevertheless harbour reservations and suggest that this result indicates a failure of the quark+diquark Faddeev equation in describing some higher baryon resonances; but resolving that question is essential in order to ensure arrival at a reliable Poincar\'e covariant description of baryon spectra and structure.  Notably, unlike the other systems described here, $\Delta(1940)\tfrac{3}{2}^-$ is only a ``$\ast\ast$'' state \cite{Workman:2022ynf}; and no electrocoupling data are currently available, so an empirical test of the wave function drawn in Fig.\,\ref{LFigures}D is not yet possible.

This discussion emphasises the role of baryon excited states in revealing novel expressions of the three pillars of EHM and the unique capacity of large momentum transfer resonance electroexcitation experiments to test theory predictions based upon them \cite{Carman:2023zke}.

\subsection{Nature's most fundamental Nambu-Goldstone boson}
This statement is made with reference to the pion \cite{Horn:2016rip}, whose manifold ``peculiarities'' can be understood from simple beginnings; namely, a collection of Goldberger-Treiman-like identities, derived and explained in Refs.\,\cite{Maris:1997hd, Qin:2014vya}, which appear as corollaries of DCSB.  These identities entail that, amongst all systems, the nearly-massless pion provides the cleanest window onto EHM.

\begin{figure}[t]
\begin{tabular}{lcl}
%
\includegraphics[clip, width=0.47\textwidth]{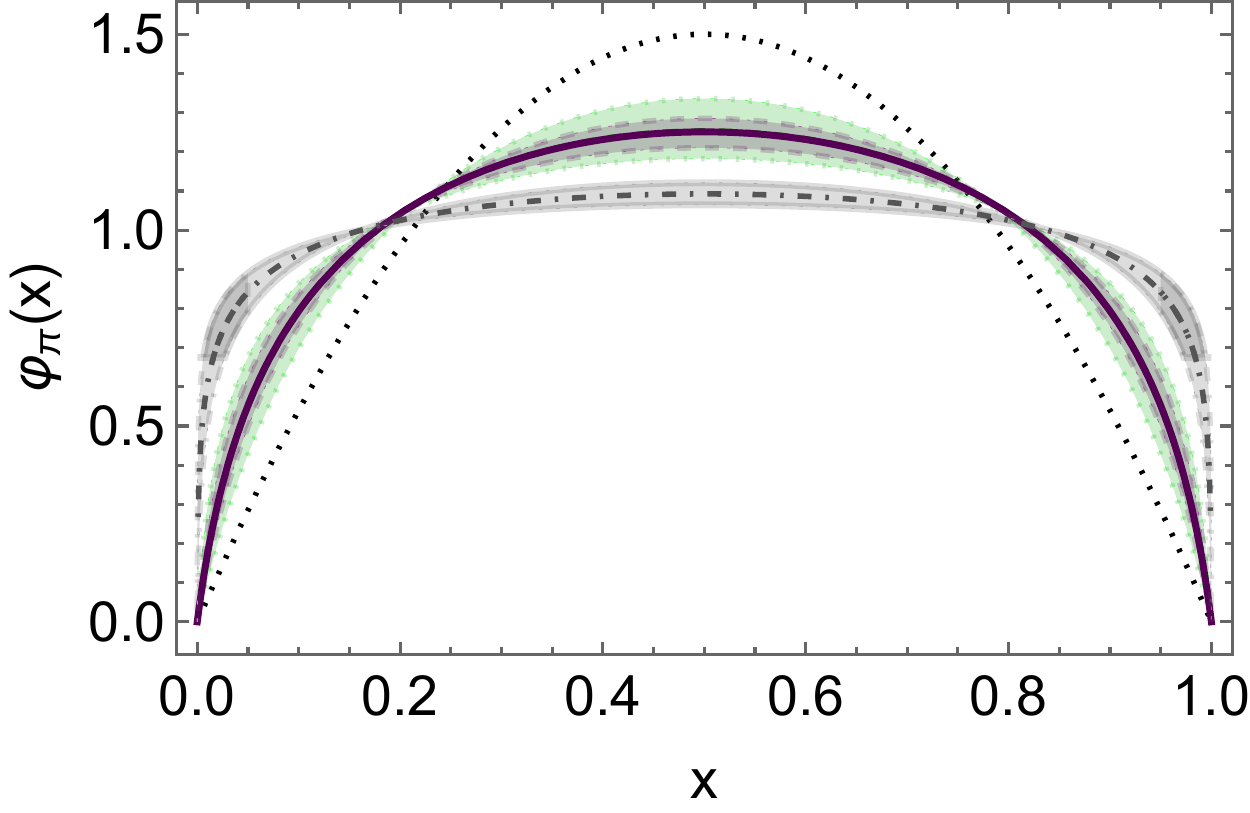} & \hspace*{-2ex} &
\includegraphics[clip, width=0.47\textwidth]{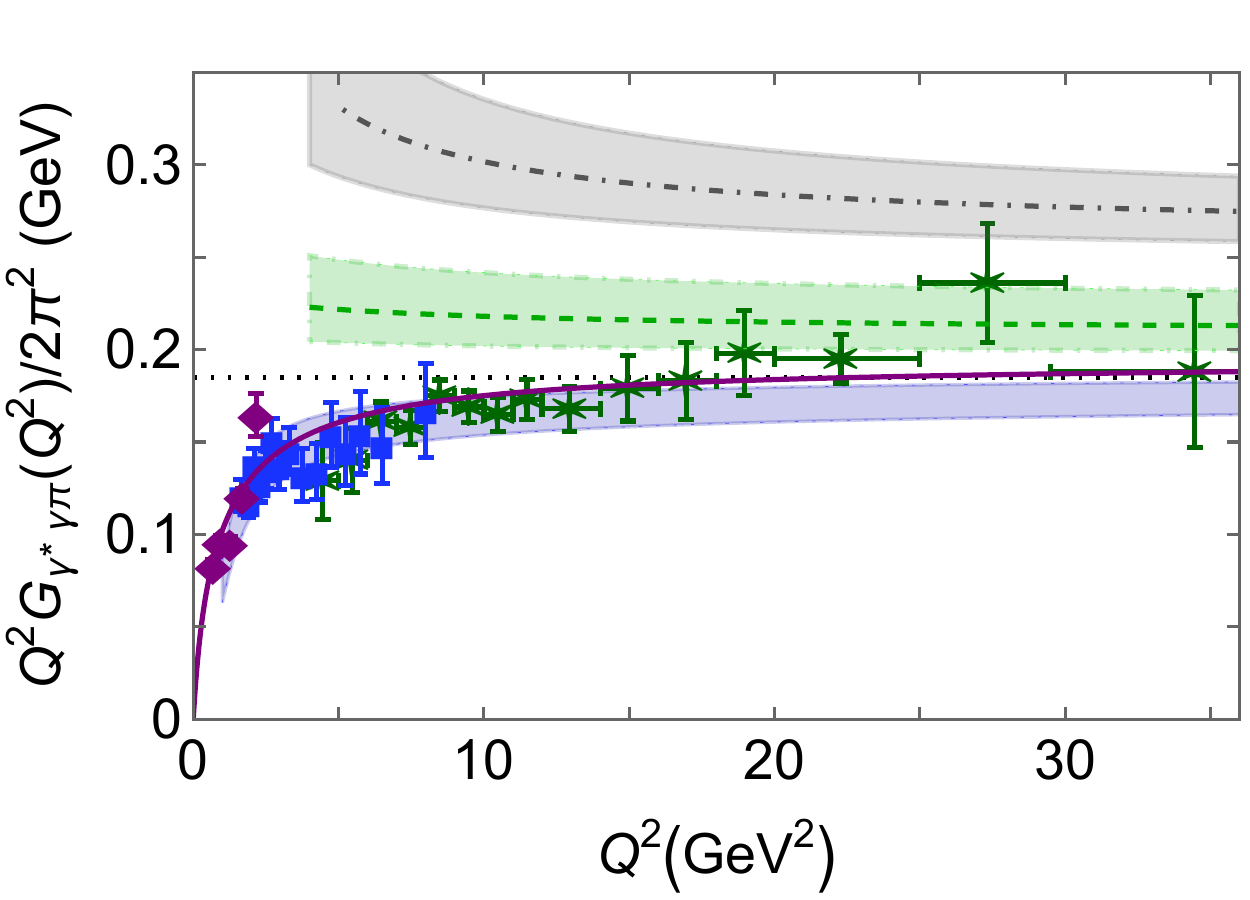}\\[-2ex]
{(\textbf{A})} & \hspace*{-2ex}  & {(\textbf{B})}
\end{tabular}
%
\caption{\label{ImagepionDA}
\textbf{A}.
Pion leading twist dressed valence quark DA.
Purple solid curve -- CSM prediction, Eq.\,\eqref{phiDB};
green dashed curve and band -- DA in form of Eq.\,\eqref{phiDB}, producing Eq.\,\eqref{z2values};
black dotted curve -- $\varphi_{\rm as}(x)=6 x(1-x)$;
grey dot-dashed curve and band -- DA in form of Eq.\,\eqref{phiDB}, producing Eq.\,\eqref{fatDA}.
\textbf{B}.
$Q^2 G_{\gamma^\ast \gamma \pi}(Q^2)/(2\pi^2)$.
Curves.
Blue band -- Ref.\,\cite{Bakulev:2012nh};
solid purple curve -- Ref.\,\cite{Raya:2015gva};
dotted black -- asymptotic limit, derived from Eq.\,\eqref{BLuv};
green dashed curve and like-coloured band -- Eq.\,\eqref{BLuv} evaluated using Eq.\,\eqref{phiDB} constrained by Eq.\,\eqref{z2values};
grey dot-dashed curve and like-coloured band -- Eq.\,\eqref{BLuv} evaluated using Eq.\,\eqref{phiDB} constrained by Eq.\,\eqref{fatDA}.
The mean-value of the ratio of the grey and green bands is $1.33(11)$.
Data:
\cite[CELLO]{Behrend:1990sr} -- diamonds (purple);
\cite[CLEO]{Gronberg:1997fj} -- squares (blue);
\cite[Belle]{Uehara:2012ag} -- stars (green).
}
\end{figure}

Significant steps toward understanding the pion were made with the CSM calculation of the pion's leading-twist parton distribution amplitude (DA) \cite{Chang:2013pq}.  In modern terms, the prediction may be expressed as follows:
\begin{equation}
\label{phiDB}
\varphi_\pi(x;\zeta_2) = {\mathpzc n}_0 \ln (1 + x (1 - x)/\rho_\varphi^2 )\,,\quad \rho_\varphi= 0.16(5)\,,
\end{equation}
with ${\mathpzc n}_0$ ensuring unit normalisation.  In comparison with the asymptotic DA \cite{Lepage:1979zb, Efremov:1979qk, Lepage:1980fj}: $\varphi_{\rm as}(x)=6 x(1-x)$, the dilation expressed in Eq.\,\eqref{phiDB} -- evident in Fig.\,\ref{ImagepionDA}A -- is a signal of EHM.  It has long been known \cite{Chernyak:1983ej} that such dilation must have a big impact on form factors that characterise hard exclusive processes.  

In the context of the pion elastic electromagnetic form factor, a detailed discussion of this fact is presented elsewhere \cite[Fig.\,4.12]{Roberts:2021nhw}.
The point can also be elaborated using the $\gamma^\ast \gamma \to \pi^0$  transition form factor, $G_{\gamma^\ast \gamma \pi}(Q^2)$, for which \cite{Lepage:1980fj}
\begin{equation}
\label{BLuv}
\exists Q_0 > m_N \, | \, Q^2 G_{\gamma^\ast \gamma \pi}(Q^2) \stackrel{Q^2 > Q_0^2}{\approx}
4 \pi^2 f_\pi \frac{1}{3}\int_0^1 dx\, \frac{1}{x} \varphi_\pi(x;Q) \,,
\end{equation}
where $f_\pi \approx 0.092\,$GeV is the pion leptonic decay constant.

Comprehensive perspectives on $G_{\gamma^\ast \gamma \pi}(Q^2)$ are provided in Refs.\,\cite{Brodsky:2011yv, Bakulev:2012nh, Raya:2015gva}.  One may conclude from these studies that
in order to deliver a result in agreement with available data on $Q^2\gtrsim m_N^2$, any internally consistent calculation must express EHM in pion structure via a leading-twist DA with significant but not extreme dilation.
(The connection between EHM and $G_{\gamma^\ast \gamma \pi}(Q^2)$ on $0\leq Q^2\lesssim m_N^2$ is discussed, e.g., in Ref.\,\cite{Holl:2005vu}.)
Combining the results in Refs.\,\cite{Bakulev:2012nh, Raya:2015gva}, one can conservatively restrict the range of empirically compatible DAs to that set for which
\begin{equation}
\label{z2values}
\langle \xi^2 \rangle_{\varphi_\pi}^{\zeta_2}=
\int_0^1 dx \, (\xi=1-2x)^2\,\varphi_\pi(x;\zeta_2) = 0.25(2)\,.
\end{equation}
In terms of Eq.\,\eqref{phiDB}, this corresponds to $\rho_\varphi = 0.16_{-0.08}^{+0.13}$ and the green band in Fig.\,\ref{ImagepionDA}A.  The predictions from Refs.\,\cite{Bakulev:2012nh, Raya:2015gva} are compared in Fig.\,\ref{ImagepionDA}B.  They overlap.

Some additional context for Eq.\,\eqref{z2values} is useful.  Exploiting the fact that the DA of a ground-state light-quark pseudoscalar meson is a concave function \cite{Li:2016dzv}, then $\tfrac{1}{5} < \langle \xi^2 \rangle_{\varphi_\pi}^{\zeta_2} < \tfrac{1}{3}$, where the lower bound is obtained using $\varphi_{\rm as}$ and the upper is produced by the most dilated DA possible, \emph{viz}.\ $\varphi_{\rm p}(x)=1$, which corresponds to a pointlike (structureless) particle.  The midpoint of this domain is $\langle \xi^2 \rangle = (\tfrac{4}{15} \approx 0.27)=:\langle \xi^2 \rangle_{\rm m}$ and the constraint in Eq.\,\eqref{z2values} corresponds to
\begin{equation}
\langle \xi^2 \rangle_{\varphi_\pi}^{\zeta_2} = 0.94(8) \langle \xi^2 \rangle_{\rm m}\,.
\end{equation}
Evidently, empirically compatible DAs lie half-way between the two extremes.

Numerous models of pion structure yield results consistent with Eq.\,\eqref{z2values} -- see, e.g., Refs.\,\cite{Brodsky:2006uqa, Choi:2007yu, Zhong:2021epq, Li:2022qul}, as do lattice-QCD computations focused specifically on the determination of this moment \cite{Braun:2015axa, RQCD:2019osh}.  When DAs with this character are used in Eq.\,\eqref{BLuv}, one obtains the green band in Fig.\,\ref{ImagepionDA}B.  The mismatch between this result, on one hand, and, on the other, the internally consistent calculations and truly asymptotic limit, Eq.\,\eqref{BLuv}, can readily be explained by higher-order (HO) corrections \cite{Li:2013xna}: compared with the asymptotic value, the relative difference on the domain displayed is just 18(6)\%.  These facts further emphasise the reliability of Eq.\,\eqref{z2values}.

In stark contrast, values of this moment obtained in lattice-QCD attempts to compute the $x$-dependence of the DA typically lead to extreme values \cite{Bali:2018spj}:
\begin{equation}
\label{fatDA}
\langle \xi^2 \rangle_{\rm lattice-DA}^{\zeta_2} = 0.30(1)\,.
\end{equation}
The analysis in Ref.\,\cite{Xu:2018eii} suggests this is an overestimate that results from a weakness of existing lattice algorithms, which produce DAs with support far outside the physical window, $0\leq x \leq 1$, thereby returning distorted (overly dilated) images.  Using the form in Eq.\,\eqref{phiDB} to express DAs with this character, one obtains the grey band of DAs drawn in Fig.\,\ref{ImagepionDA}A.  Inserting those DAs into Eq.\,\eqref{BLuv}, one obtains the grey band in Fig.\,\ref{ImagepionDA}B.  Comparing this body of curves with existing data and other theory, one sees a mismatch that cannot be explained by HO corrections:  compared with the asymptotic value, the relative difference on the domain displayed is 59(13)\%.  Consequently, DAs characterised by Eq.\,\eqref{fatDA} are unlikely to present a realistic picture of pion structure.

\subsection{Proton and pion distribution functions}
\label{SecDFs}
In another recent advance, CSMs have been used to deliver a unified set of predictions for pion and proton parton distribution functions (DFs) \cite{Lu:2022cjx}: valence, glue, and four-flavour sea.  This has made it possible to address issues relating to the expression of EHM in the distribution of partons within two very different forms of hadron matter: the spin-half proton -- Nature's most fundamental bound state; and the spin-zero pion -- Nature's most fundamental (near) Nambu--Goldstone boson.  These studies may also be seen against a backdrop of longstanding QCD predictions.  Namely, at the hadron scale, $\zeta_{\cal H}<m_p$, valence quark DFs in the proton and pion behave as follows \cite{Brodsky:1994kg, Yuan:2003fs, Cui:2021mom, Cui:2022bxn, Chang:2022jri}:
\begin{equation}
\label{LargeX}
{\mathpzc d}^p(x;\zeta_{\cal H}), {\mathpzc u}^p(x;\zeta_{\cal H}) \stackrel{x\simeq 1}{\propto} (1-x)^3\,,
\quad
\bar {\mathpzc d}^\pi(x;\zeta_{\cal H}), {\mathpzc u}^\pi(x;\zeta_{\cal H}) \stackrel{x\simeq 1}{\propto} (1-x)^2\,.
\end{equation}
Furthermore, the DGLAP evolution equations \cite{Dokshitzer:1977sg, Gribov:1971zn, Lipatov:1974qm, Altarelli:1977zs} entail that the large-$x$ exponent on the kindred gluon DF is approximately one unit larger; and that for sea quark distributions is roughly two units larger.

\begin{figure}[t]
\begin{tabular}{lcl}
%
\includegraphics[clip, width=0.47\textwidth]{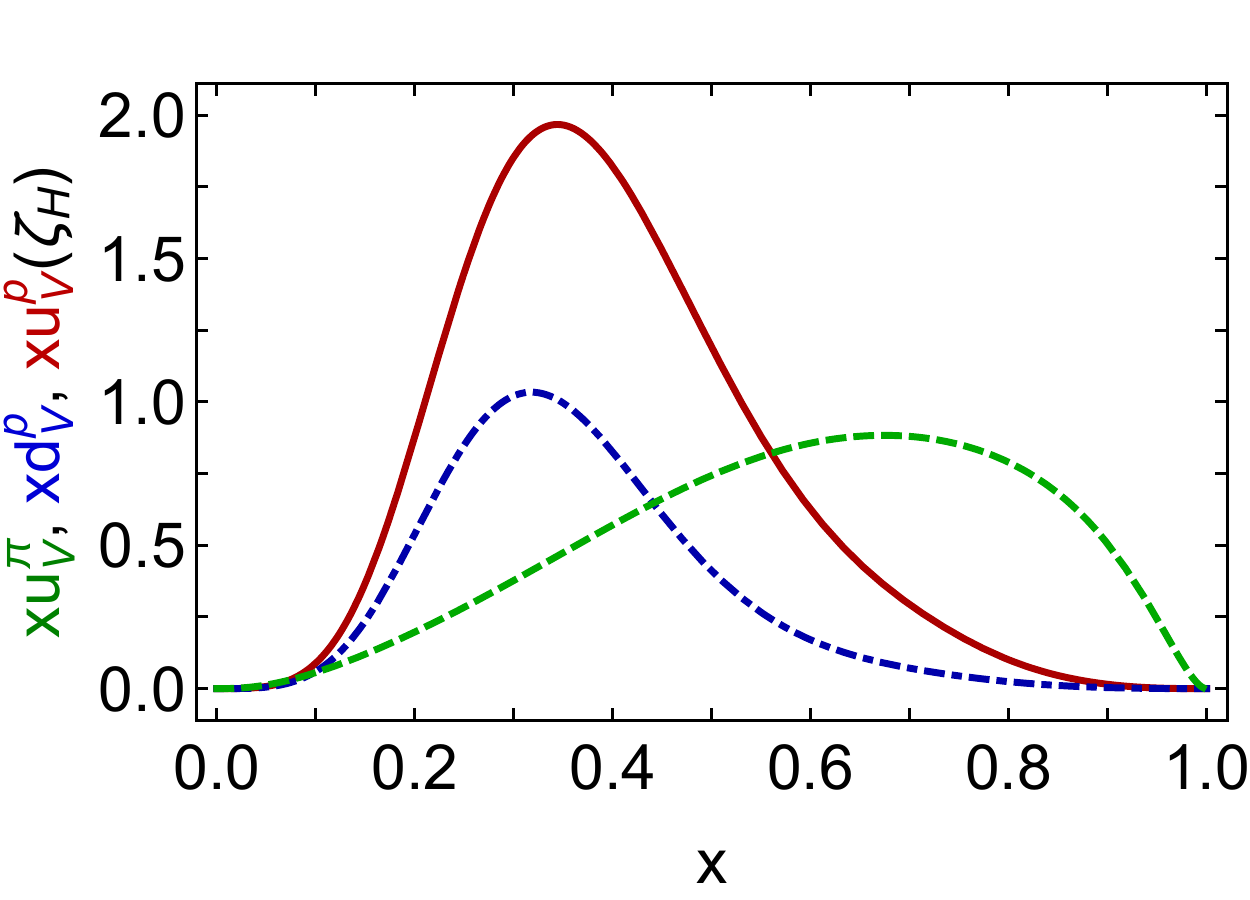} & \hspace*{-2ex} &
\includegraphics[clip, width=0.47\textwidth]{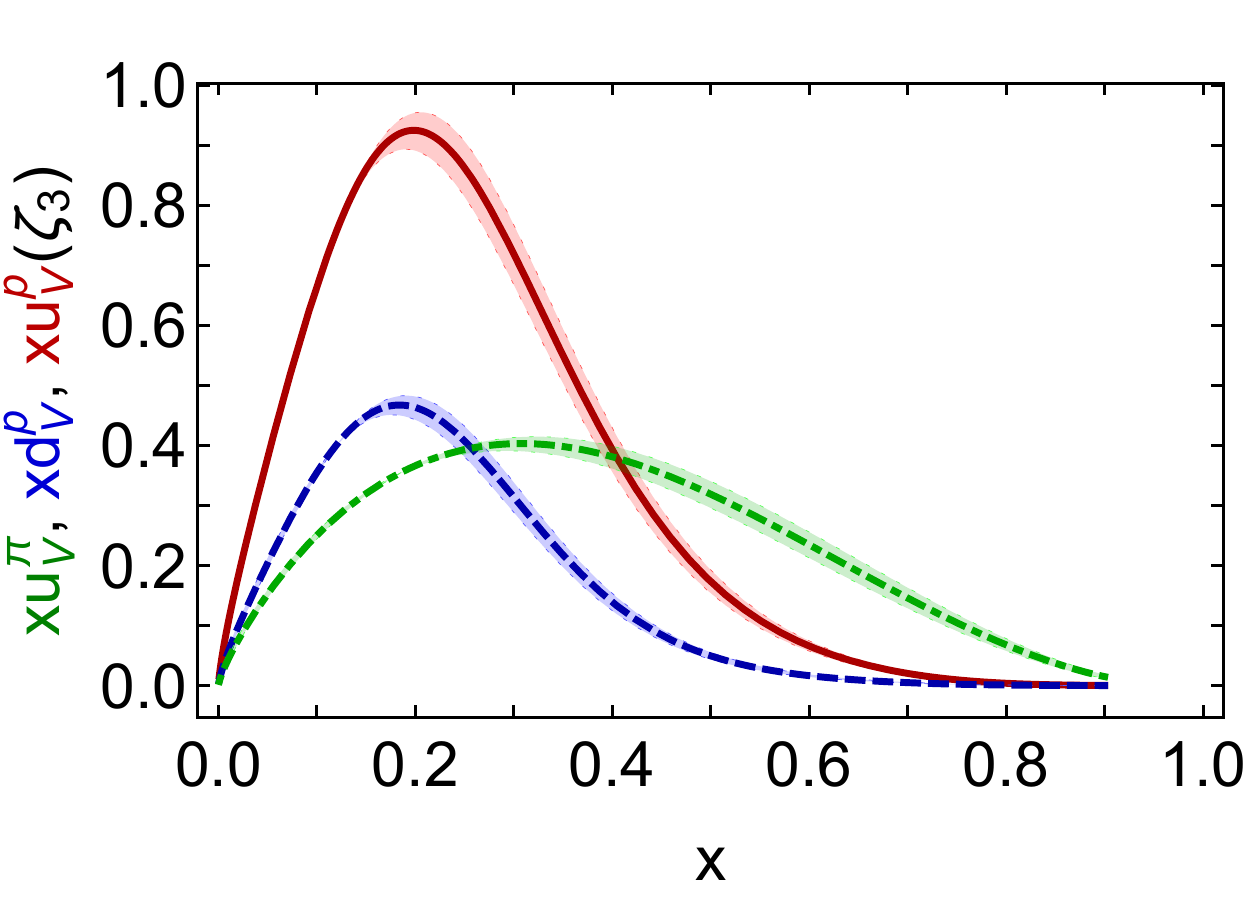}\\[-4ex]
{(\textbf{A})} &\hspace*{-2ex}  & {(\textbf{B})}\\[-3ex]
\end{tabular}
\vspace*{3ex}
\caption{\label{ImageValence}
\textbf{A}.
Hadron scale proton and pion valence parton DFs:
$x {\mathpzc u}^p(x;\zeta_{\cal H})$ -- solid red curve;
$x {\mathpzc d}^p(x;\zeta_{\cal H})$ -- dotted-dashed blue curve;
$x {\mathpzc u}^\pi(x;\zeta_{\cal H})$ -- dashed green curve.
\textbf{B}.
Valence DFs in Panel \mbox{\bf A} evolved to $\zeta_3=m_{J/\psi}=3.097\,$GeV.
The band surrounding each curve expresses the response to a $\pm 5$\% variation in $\zeta_{\cal H}$.
}
\end{figure}

The predictions from Ref.\,\cite{Lu:2022cjx} are displayed in Fig.\,\ref{ImageValence}.  The following points are worth highlighting.
(\emph{a}) Each DF is consistent with the relevant entry in Eq.\,\eqref{LargeX}.
(\emph{b}) At $\zeta_{\cal H}$, as noted above, the momentum sum rules for each hadron are saturated by valence degrees-of-freedom, \emph{viz}.\
\begin{equation}
\langle x \rangle_{{\mathpzc u}_p}^{\zeta_{\cal H}}=0.687\,,\;
\langle x \rangle_{{\mathpzc d}_p}^{\zeta_{\cal H}} = 0.313\,,\;
\langle x \rangle_{{\mathpzc u}_\pi}^{\zeta_{\cal H}} =0.5 = \langle x \rangle_{\bar{\mathpzc d}_\pi}^{\zeta_{\cal H}}\,.
\end{equation}
(\emph{c}) The presence of diquark correlations in the proton entails that
$\langle x \rangle_{{\mathpzc u}_p}^{\zeta_{\cal H}} > 2 \langle x \rangle_{{\mathpzc d}_p}^{\zeta_{\cal H}}$.
(\emph{d}) Proton and pion DFs have markedly different $x$-dependence, with the pion valence-quark DFs being very much more dilated than those of the proton.  In part, this is a consequence of Eqs.\,\eqref{LargeX}.  However, it also owes to EHM: pion DFs are the most dilated in Nature.  This feature is seen, too, in the symmetry-preserving treatment of a vector$\times$vector contact interaction \cite{YangYuPrivate}.  It is preserved under evolution -- see Fig.\,\ref{ImageValence}B.

\begin{figure}[t]
\begin{tabular}{lcl}
%
\includegraphics[clip, width=0.47\textwidth]{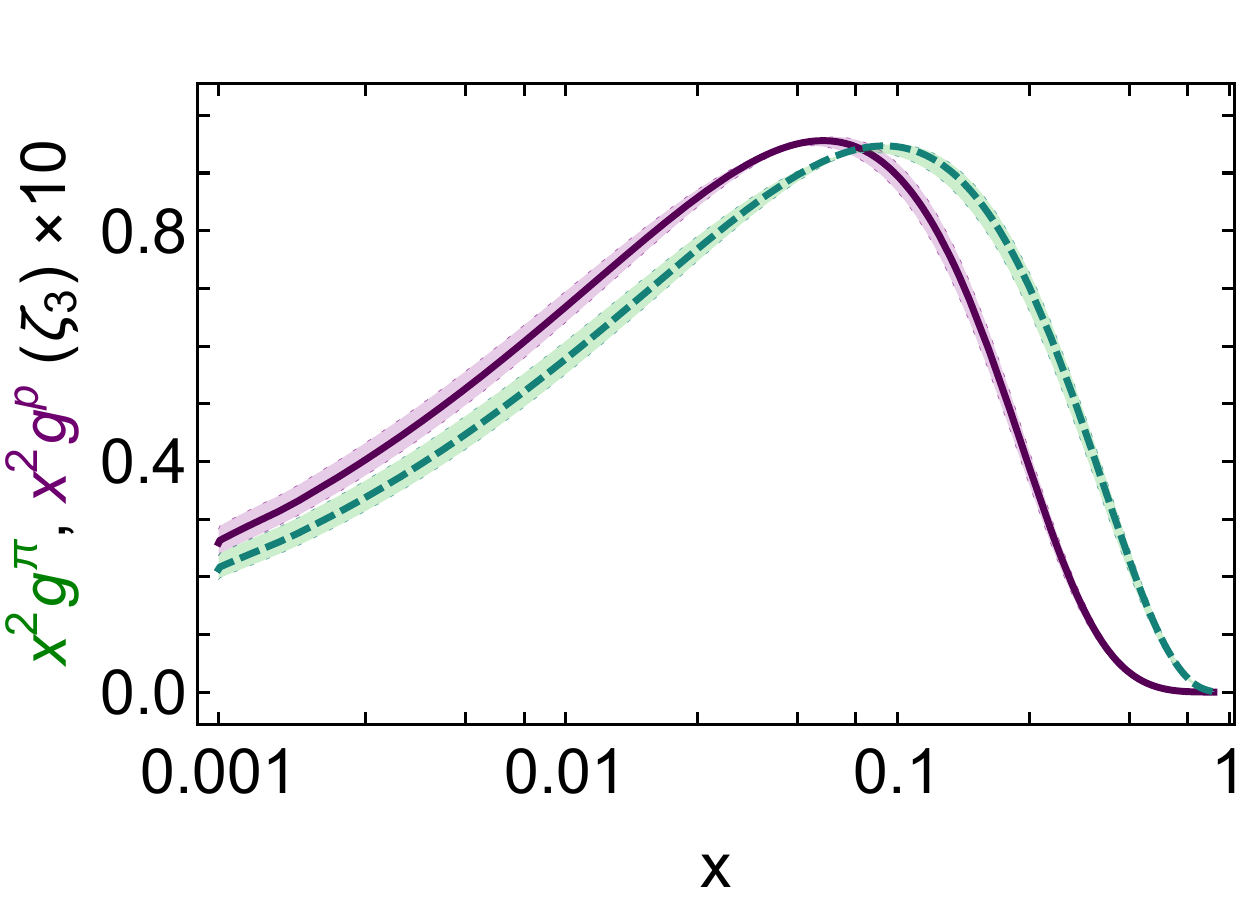} & \hspace*{-2ex} &
\includegraphics[clip, width=0.47\textwidth]{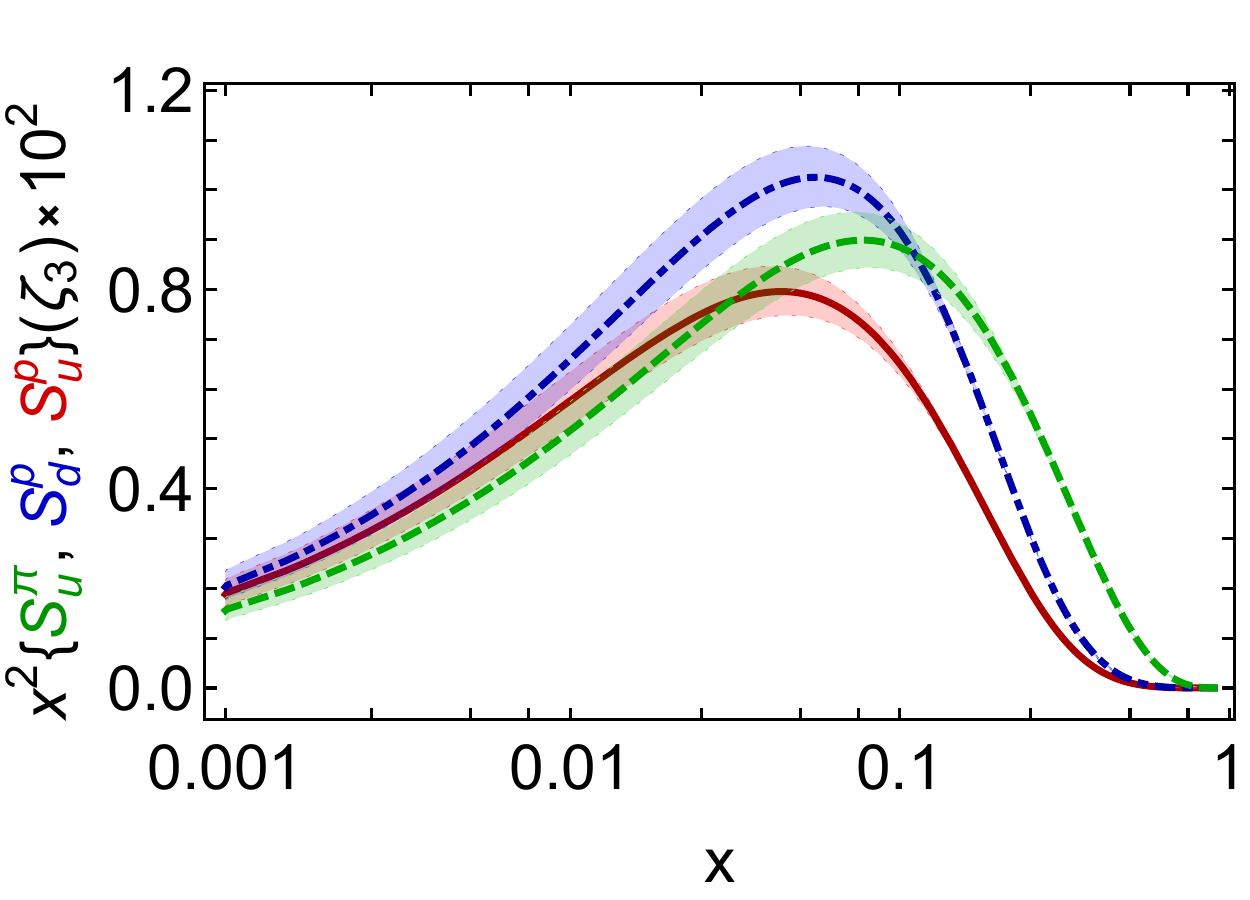}\\[-2ex]
{(\textbf{A})} & \hspace*{-2ex}  & {(\textbf{B})}
\end{tabular}
%
\caption{\label{ImageGlue}
\textbf{A}.
Glue DFs: $x^2 {\mathpzc g}$, in the proton (solid purple curve) and pion (dashed green curve) at $\zeta=\zeta_3$. 
\textbf{B}.
Proton and pion light quark sea DFs:
$x^2 {\mathpzc S}_u^p(x;\zeta_{3})$---solid red curve;
$x^2 {\mathpzc S}_d^p(x;\zeta_{3})$---dashed blue curve;
$x^2 {\mathpzc S}_u^\pi(x;\zeta_{3})$---dotted-dashed green curve.
The band surrounding each curve expresses the response to a $\pm 5$\% variation in $\zeta_{\cal H}$.
}
\end{figure}

CSM predictions for $\zeta=\zeta_3=3.1\,$GeV proton and pion glue DFs are drawn in Fig.\,\ref{ImageGlue}A.  They are consistent with the remarks following Eq.\,\eqref{LargeX}.
As discussed elsewhere \cite{Chang:2021utv}, the glue-in-$\pi$ DF agrees with that obtained in a recent lattice-QCD computation \cite{Fan:2021bcr}.
Furthermore, reproducing the EHM-induced pattern seen with valence quark DFs in Figs.\,\ref{ImageValence}, Fig.\,\ref{ImageGlue}A reveals that the glue-in-$\pi$ DF possesses significantly more support on the valence domain, $x\gtrsim 0.1$, than the glue-in-$p$ DF.

The $\zeta=\zeta_3$ proton and pion light-quark sea DFs are drawn in Fig.\,\ref{ImageGlue}B. The EHM-driven pattern of dilation is also apparent here, with the sea-in-$\pi$ DF possessing greater support on $x\gtrsim 0.1$ than the kindred sea-in-$p$ DFs.

In the CSM study \cite{Lu:2022cjx}, DFs of the heavier sea quarks are also generated via evolution.  The $\zeta=\zeta_3$ $s$ and $c$ quark DFs are similar in size to those of the light quark sea DFs; and for these heavier quarks, too, the pion DFs have greater support on the valence domain, $x\gtrsim 0.1$, than the related proton DFs.

All sea-quark DFs are consistent with the remarks following Eq.\,\eqref{LargeX}, i.e., have $x$-profiles typical of sea DFs.

Notably, in both the pion and proton, the CSM study predicts $\langle x\rangle_{{\mathpzc c}}^{\zeta=1.5\,{\rm GeV}}=0.64(3)$\%.  Nothing is known about this momentum fraction in the pion; and in the proton, phenomenological estimates are vague, ranging from $0$-$2$\% \cite[Fig.\,59]{NNPDF:2017mvq}.  Nonetheless, the analysis in Ref.\,\cite{Lu:2022cjx} predicts a significant $c$ quark momentum fraction without recourse to ``intrinsic charm'' \cite{Brodsky:1980pb}; hence, are something of a challenge to the claims in Ref.\,\cite{Ball:2022qks}.
%

It is worth closing this subsection by stressing that DF dilation is a ``smoking gun'' for EHM.

\subsection{MARATHON}
The ratio of neutron and proton structure functions has long been recognised as a keen differentiator between pictures of nucleon structure \cite{Roberts:2013mja, CLAS:2014jvt, Abrams:2021xum}: on the valence-quark domain,
$F_2^{n}(x)/F_2^{p}(x) = [1 + 4 r(x)]/[4+r(x)]$, $r(x) = {\mathpzc d}(x)/{\mathpzc u}(x)$.
The problem in obtaining an empirical result for $F_2^{n}(x)/F_2^{p}(x)$ is the measurement of $F_2^{n}$ because isolated neutrons decay fairly quickly.  Motivated by Refs.\,\cite{Bodek:1973dy, Poucher:1973rg}, many experiments have used the deuteron; but despite being weakly bound, the representation dependence of proton-neutron interactions in the deuteron leads to big theory uncertainties in $F_2^{n}(x)/F_2^{p}(x)$ on $x\gtrsim 0.7$ \cite{Whitlow:1991uw}.

\begin{figure}[t]
\begin{tabular}{lcl}
%
\includegraphics[clip, width=0.47\textwidth]{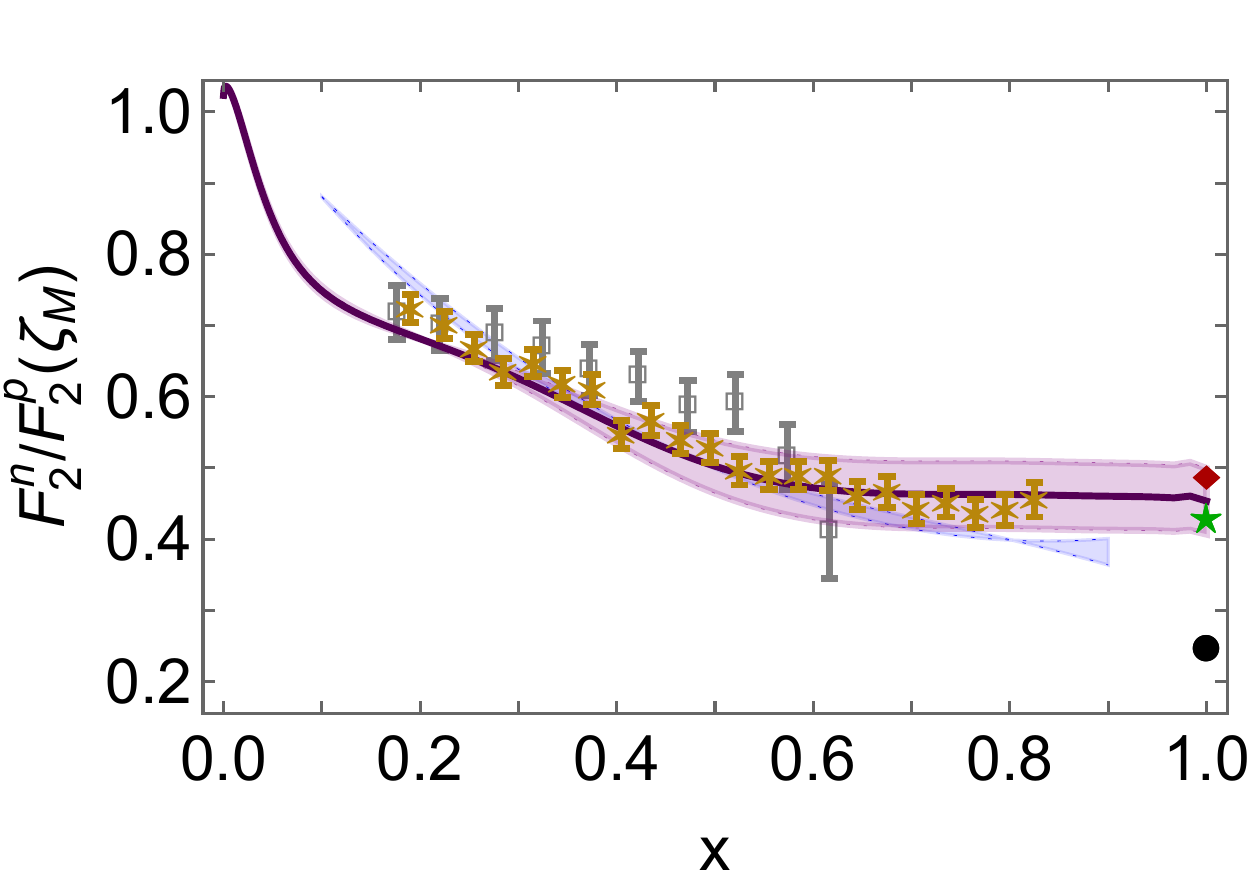} & \hspace*{-2ex} &
\includegraphics[clip, width=0.47\textwidth]{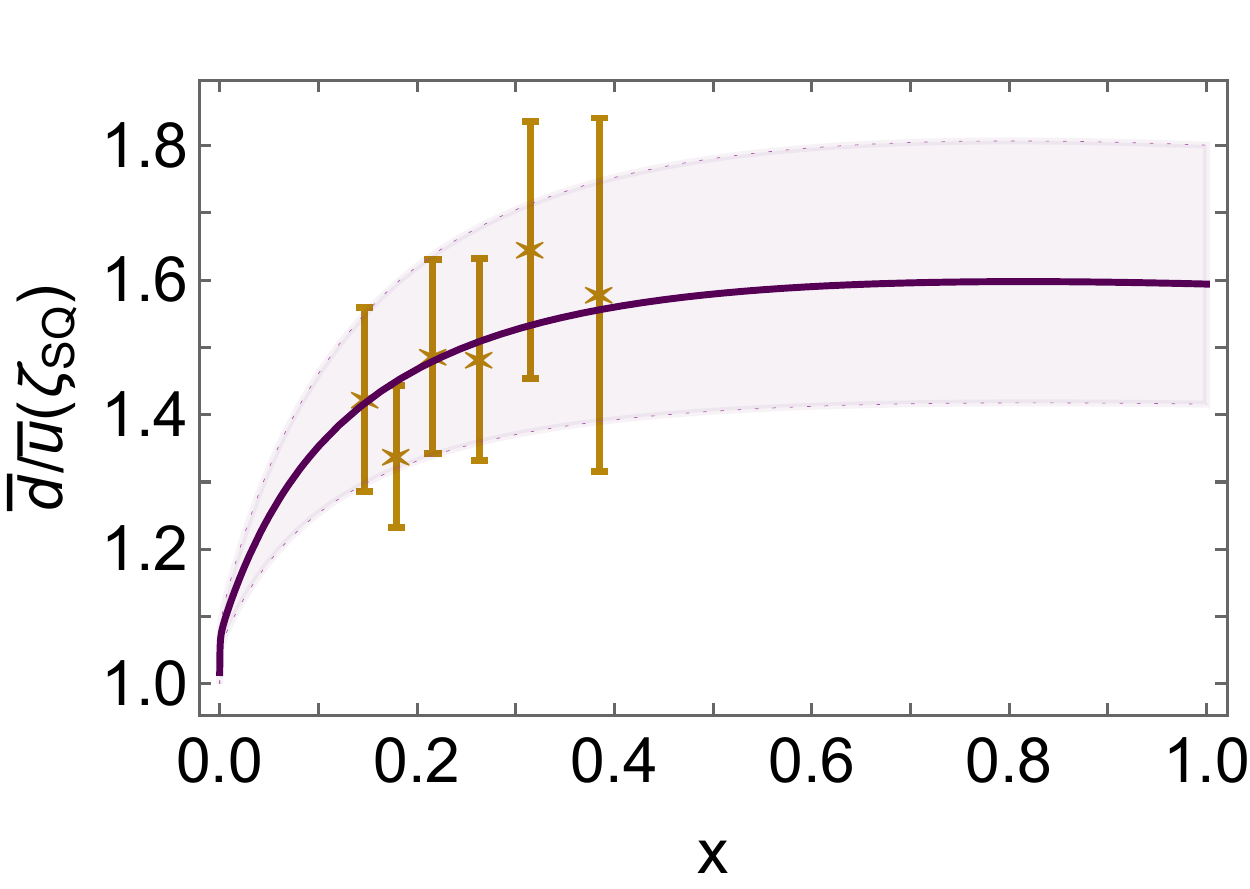}\\[-3ex]
{(\textbf{A})} & \hspace*{-2ex} & {(\textbf{B})}
\end{tabular}
%
\caption{\label{ImageSeaQuest}
\textbf{A}.
Neutron:proton structure function ratio.
Data: \cite[BoNuS]{CLAS:2014jvt} -- open grey squares;
\cite[MARATHON]{Abrams:2021xum} -- gold asterisks.
Light-blue band  -- alternate analysis of MARATHON data using light-front Hamiltonian dynamics \cite{Pace:2022qoj}.  This band is only well constrained within the domain covered by the MARATHON data.  On $x>0.7$, it expresses the large uncertainties that exist thereupon in data-based inferences of the proton structure function \cite{Amoroso:2022eow}.
CSM prediction (solid purple curve) obtained from valence quark DFs in Fig.\,\ref{ImageValence} after evolution to $\zeta=\zeta_{M}=2.7\,$GeV \cite{Chang:2022jri, Lu:2022cjx}.
Other predictions:
green star -- helicity conservation in the parton model \cite{Farrar:1975yb, Brodsky:1979gy};
red diamond -- large-$x$ estimate based on Faddeev equation solutions \cite{Roberts:2013mja};
and filled circle -- scalar-diquark-only proton wave function \cite{Xu:2015kta}.
The band bracketing the CSM curve (only noticeable on the valence quark domain) expresses the response to a $\pm 5$\% variation in the strength of axialvector diquark contributions to the proton charge.
\textbf{B}.
Ratio of light antiquark DFs.
Data: Ref.\,\cite[E906]{SeaQuest:2021zxb}.
CSM result (solid purple curve) obtained from valence quark DFs in Fig.\,\ref{ImageValence} after evolution to $\zeta^2=\zeta_{\rm SQ}^2 = 30\,$GeV$^2$ \cite{Lu:2022cjx}. The shaded band expresses the impact of a $\pm 25$\% variation in the strength of Pauli blocking.
}
\end{figure}

A better approach is provided by comparisons of deep inelastic scattering (DIS) from $^3$H and $^3$He because nuclear interaction effects cancel to a very large extent when extracting $F_2^n(x)/F_2^p(x)$ from the $^3$H:$^3$He scattering rate ratio \cite{Afnan:2000uh, Pace:2001cm}.  Here the problem is that $^3$H is highly radioactive; so, great care is required to deliver a safe target.  Recently, following many years of development, the obstacles were overcome and such an experiment completed \cite{Abrams:2021xum}.  The data are reproduced in Fig.\,\ref{ImageSeaQuest}A.  
Significantly, within mutual uncertainties, the data from Ref.\,\cite{Abrams:2021xum} match results inferred from analyses of nuclear DIS reactions that account for the effects of short-range correlations in nuclei \cite{Segarra:2019gbp}.

Recalling Sec.\,\ref{SecFE}, CSMs make clear statements about proton structure; in particular, axialvector diquark correlations are responsible for approximately 40\% of the proton's charge.  This structural picture is confirmed, e.g., by the agreement between data and associated predictions for nucleon axial structure -- see Sec.\,\ref{SecAxial}; and it leads to the proton valence-quark DFs discussed in Sec.\,\ref{SecDFs}.  Using those results, a prediction for the neutron-proton structure function ratio follows immediately:
\begin{align}
\label{F2nF2p}
\frac{F_2^n(x;\zeta)}{F_2^p(x;\zeta)} =
\frac{
{\mathpzc U}(x;\zeta) + 4 {\mathpzc D}(x;\zeta) + \Sigma(x;\zeta)}
{4{\mathpzc U}(x;\zeta) + {\mathpzc D}(x;\zeta) + \Sigma(x;\zeta)}\,,
\end{align}
where, in terms of quark and antiquark DFs,
${\mathpzc U}(x;\zeta) = {\mathpzc u}^p(x;\zeta)+\bar {\mathpzc u}^p(x;\zeta)$,
${\mathpzc D}(x;\zeta) = {\mathpzc d}^p(x;\zeta)+\bar {\mathpzc d}^p(x;\zeta)$, and
$\Sigma(x;\zeta) = {\mathpzc s}^p(x;\zeta)+\bar {\mathpzc s}^p(x;\zeta)
 +{\mathpzc c}^p(x;\zeta)+\bar {\mathpzc c}^p(x;\zeta)$.
The $\zeta=2.7\,{\rm GeV}=:\zeta_M$ CSM prediction is drawn in Fig.\,\ref{ImageSeaQuest}A.  In comparison with modern data \cite[MARATHON]{Abrams:2021xum}, the central CSM curve yields $\chi^2/$ degree-of-freedom$\;=1.3$.
Notably, the $x$-dependence of the CSM prediction was made without reference to any data.  Consequently, the agreement with MARATHON data is meaningful.
Moreover, the CSM $x\simeq 1$ value matches that determined elsewhere \cite{Cui:2021gzg} using a robust
method for extrapolating the MARATHON data to arrive at a model-independent result.
Once again, these things support the CSM prediction that axialvector diquark correlations contribute significantly to nucleon structure and cannot reasonably be ignored.

\subsection{Antimatter asymmetry in the proton}
In common implementations of DGLAP evolution, gluon splitting produces quark+antiquark pairs of all flavours with equal probability.  Physically, however, given that the proton contains two valence $u$ quarks and one valence $d$ quark, the Pauli exclusion principle should force gluon splitting to prefer $d+\bar d$ production over $u+\bar u$ \cite{Field:1976ve}.  This effect was implemented in Refs.\,\cite{Chang:2022jri, Lu:2022cjx} by introducing a small Pauli blocking factor into the gluon splitting function and led therein to the difference between proton light-quark sea DFs evident in Fig.\,\ref{ImageGlue}B.

Such a difference between the light-quark sea DFs entails a violation of the Gottfried sum rule \cite{Gottfried:1967kk, Brock:1993sz}, which has been seen \cite{NewMuon:1991hlj, NewMuon:1993oys, NA51:1994xrz, NuSea:2001idv, SeaQuest:2021zxb}.  On the domain covered by the measurements in Refs.\,\cite{NewMuon:1991hlj, NewMuon:1993oys}, the proton DFs in Fig.\,\ref{ImageGlue}B yield
\begin{equation}
\label{gottfried}
\int_{0.004}^{0.8} dx\,[\bar {\mathpzc d}(x;\zeta_3) - \bar {\mathpzc u}(x;\zeta_3)]
= 0.116(12)\,.
\end{equation}
This value for the Gottfried sum rule discrepancy matches that inferred from recent fits to a large sample of high-quality data ($\zeta = 2\,$GeV) 0.110(80) \cite[CT18]{Hou:2019efy}, and is far more precise.
%
Regarding the strength of Pauli blocking, the result in Eq.\,\eqref{gottfried} equates with a term in the gluon splitting function that shifts just $\approx 25$\% of the $u$ quark sea momentum fraction into the $d$ quark sea at $\zeta = \zeta_2$.  The indicated uncertainty expresses the impact of a $\pm 25$\% change in that strength.

Data from the most recent experiment to explore the asymmetry of antimatter in the proton \cite[E906]{SeaQuest:2021zxb} are displayed in Fig.\,\ref{ImageSeaQuest}B and compared therein with the CSM result obtained using the proton DFs in Fig.\,\ref{ImageGlue}B. 
Plainly, to explain modern data, a modest Paul blocking effect in the gluon splitting function is sufficient.

Pursuing these notions further, at resolving scales typical of measurements that are interpretable in terms of DFs, the $s$-quark sea in the pion should be larger than either of its light-quark sea components.  Calculations are underway to quantify this effect.

\subsection{Distribution of mass in the proton}
Before closing, it is worth remarking here that a new approach to the analysis of existing data on pion valence quark DFs \cite{Conway:1989fs} and the pion elastic electromagnetic form factor \cite{Amendolia:1984nz, Amendolia:1986wj, Volmer:2000ek, Horn:2006tm, Tadevosyan:2007yd, Blok:2008jy, Huber:2008id} has been used \cite{Xu:2023bwv} to develop data-based predictions for the three-dimensional structure of the pion and, therefrom, the pion mass distribution form factor, $\theta_2^\pi$.  Measured against the pion elastic electromagnetic (em) form factor, $\theta_2^\pi$ is harder: the ratio of radii is $r_\pi^{\theta_2}/r_\pi^{\rm em} = 0.79(3)$.  Evidently, the pion mass distribution is far more compact than its charge distribution.  In fact, the pion's mass is localized within a spacetime volume that is only 40\% of the domain occupied by its charge.

At first sight, these results may appear surprising.  However, they may readily be understood.  The pion's Bethe-Salpeter wave function is a property of the pion.  It is independent of the probing object; so, is the same whether a photon or graviton is used.  However, the probe itself reacts to different features of the target constituents.  A target quark carries the same charge, independent of its momentum.  So, the wave function alone controls the distribution of charge.  On the other hand, the gravitational interaction of a target quark depends on its momentum; obviously, because the current relates to the energy-momentum tensor.  The pion mass distribution therefore depends on interference between quark momentum growth and wave function fall-off.  This drives support to a larger momentum domain within the pion, which corresponds to a smaller distance domain.

It is further argued in Ref.\,\cite{Xu:2023bwv} that one may place the following tight constraints on the ratio of radii:
\begin{equation}
\label{bounds}
\tfrac{1}{\surd 2} \leq r_\pi^{\theta_2} /r_\pi^{\rm em} \leq 1\,,
\end{equation}
where the lower bound is saturated by a point-particle DF and the upper by the DF of a bound-state formed from infinitely massive valence degrees-of-freedom.

Arguing by analogy, the distribution of mass within the proton is very probably also more compact than the distribution of charge.  It should be possible to generalise the methods in Ref.\,\cite{Xu:2023bwv} and quantify this observation.  Such a comparison could potentially reveal much about the character of EHM.

\section{Perspective}\label{sec13}
In the six years that have elapsed since the previous Baryons meeting, significant progress has been made in the study of hadron structure using continuum Schwinger function methods (CSMs), only a few highlights of which have been sketched above.  Additional contributions are discussed elsewhere \cite{Roberts:2020udq, Roberts:2020hiw, Roberts:2021xnz, Roberts:2021nhw, Binosi:2022djx, Roberts:2022rxm, Papavassiliou:2022wrb, Ding:2022ows, Ferreira:2023fva, Carman:2023zke}.  Arguably the most important steps were made with identification of the three pillars of emergent hadron mass (EHM) in the Standard Model: the running gluon mass, process-independent effective charge, and running quark mass.  These pillars support the position that QCD is the first well-defined four-dimensional quantum field theory that has ever been contemplated.

There are now good reasons to adopt a perspective from which QCD appears unique amongst known fundamental theories of natural phenomena.
The degrees-of-freedom that are used to express the scale-free QCD Lagrangian density are not directly observable.  One may question whether they are anything more than a choice of variables useful in developing a perturbation theory.
Notwithstanding that, the theory expresses remarkable dynamical features.
(\emph{i}) The Lagrangian's massless gluon partons are transmogrified into massive quasiparticle gauge bosons, with no ``human intervention''; namely, without recourse to any Higgs-like mechanism.  Gluon self-interactions are necessary and sufficient to dynamically generate the gluon mass function.
(\emph{ii}) The emergence of the gluon mass ensures a stable, infrared completion of QCD through appearance of a running coupling that saturates at infrared momenta, being everywhere finite.
(\emph{iii}) Together, via a matter sector gap equation,  (\emph{i}) and (\emph{ii}) ensure that massless quark partons become massive.
That, in turn, absent Higgs boson couplings into QCD, leads simultaneously to a collection of massless flavour nonsinglet pseudoscalar Nambu-Goldstone bosons and massive hadrons in all other channels.

EHM is expressed in every strong interaction observable and differently in each system.  Turning on Higgs boson couplings, it can also be revealed in processes that are sensitive to EHM+Higgs boson interference, such as weak-interaction transition form factors \cite{Xu:2021iwv, Xing:2022sor, Yao:2021pyf, Yao:2021pdy}.

High-luminosity, high-energy facilities are the key to providing science with the diverse array of hadron ``targets'' that is necessary in order to test this EHM paradigm.  The potential reward is great because validation may provide insights that enable extension of the Standard Model using only those four spacetime dimensions with which are now familiar.

\backmatter


\bmhead{Acknowledgments}
This contribution is based on results obtained and insights developed through collaborations with many people, to all of whom I am greatly indebted.
Work supported by
National Natural Science Foundation of China (grant no.\,12135007).


\end{document}